\documentclass[12pt,iop,letter]{emulateapj}
\usepackage{upgreek}
\RequirePackage{lineno}

\shortauthors{KERR ET AL.}
\shorttitle{Fermi Pulsar Timing}

\begin{document}

%
\def\ergs{\,erg\,s$^{-1}$}
\def\ergcmsqs{\,erg\,\,cm$^{-2}$\,s$^{-1}$}
\def\pccc{pc\,cm$^{-3}$}
\def\g{$\gamma$}
\def\pasa{PASA}
\def\fm{\textit{Fermi}}
\def\nudot{$\dot{\nu}$}
\def\nudotdot{$\ddot{\nu}$}


\def\hr{^{\rm h}}
\def\mi{^{\rm m}}
\def\msun{M_{\sun}}
\def\todo{\textbf{TODO} }
\def\nump{81}
\def\numrq{37}

\title{Timing Gamma-ray Pulsars with the \textit{Fermi} Large Area
Telescope: Timing Noise and Astrometry}
\author{
M.~Kerr\altaffilmark{1,2}, 
P.~S.~Ray\altaffilmark{3},
S.~Johnston\altaffilmark{1},
R.~M.~Shannon\altaffilmark{1},
and F.~Camilo\altaffilmark{1}
}
\altaffiltext{1}{CSIRO Astronomy and Space Science, Australia Telescope National
\altaffiltext{2}{email: matthew.kerr@gmail.com}
Facility, PO~Box~76, Epping NSW~1710, Australia}
\altaffiltext{3}{Space Science Division, Naval Research Laboratory, Washington, DC 20375-5352, USA}
\altaffiltext{4}{Columbia Astrophysics Laboratory, Columbia University,
  New York, NY~10027, USA}

\begin{abstract}
We have constructed timing solutions for \nump{} \g-ray pulsars
covering more than five years of \fm{} data.  The sample includes
\numrq{} radio-quiet or radio-faint pulsars which cannot be timed with
other telescopes.  These timing solutions and the corresponding pulse
times of arrival are prerequisites for further study, e.g.
phase-resolved spectroscopy or searches for mode switches.  Many
\g-ray pulsars are strongly affected by timing noise, and we present a
new method for characterizing the noise process and mitigating its
effects on other facets of the timing model.   We present an analysis
of timing noise over the population using a new metric for
characterizing its strength and spectral shape, namely its time-domain
correlation.  The dependence of the strength on $\nu$ and \nudot{} is
in good agreement with previous studies.  We find that noise process
power spectra $S(f)$ for unrecycled pulsars are steep, with strong
correlations over our entire data set and spectral indices
$S(f)\propto f^{-\alpha}$ of $\alpha\sim$5--9.  One possible
explanation for these results is the occurrence of unmodelled,
episodic `microglitches'.  Finally, we show that our treatment of
timing noise results in robust parameter estimation, and in particular
we measure a precise timing position for each pulsar.  We extensively
validate our results with multi-wavelength astrometry, and using our
updated position, we firmly identify the X-ray counterpart of
PSR~J1418$-$6058.
\end{abstract}

\section{Introduction} \label{sec:intro} 

The Large Area Telescope \citep[LAT,][]{Atwood09} of the \textit{Fermi
Gamma-ray Space Telescope} has now detected more than 150 pulsars,
extending over six orders of magnitude in spin-down luminosity and
divided roughly evenly between young `normal' pulsars and `recycled'
millisecond pulsars (MSPs).   The `normal' population is further
roughly equally divided into radio-loud and radio-quiet/radio-faint
pulsars.  See the 2nd \fm{} Pulsar Catalog \citep[2PC; ][]{Abdo13_2pc}
for a detailed analysis.  This bonanza both probes the high-energy
emission mechanism and offers selection effects appreciably different
to those of radio surveys.  E.g., the sensitivity of `blind
searches'---efforts to detect $\gamma$-ray pulsations without
\textit{a priori} information about the period---for young pulsars
suffers in the bright diffuse background of the Galactic plane
\citep{Dormody11}, while \fm{} is particularly good at selecting MSP
candidates whose pulsations can be detected with deep radio telescopes
searches \citep[e.g.][]{Ransom11,Kerr12,Camilo12b}.

Important properties of pulsars, such as their frequency $\nu$, their
spin-down rate \nudot{}, their position and proper motion, and their
orbital parameters, can be determined through pulsar timing.  This
procedure---as old as pulsars themselves \citep{Hewish68}---both
measures pulse times of arrival (TOAs) and models their generation and
propagation so as to coherently account for every rotation of the
neutron star.  Measuring these parameters is key to understanding the
distribution of pulsars within the Galaxy.  Moreover, deviations from
expectations often yield new insights, e.g. pulse nulling over very
long time-scales \citep[intermittency;][]{Kramer06,Camilo12}, correlated shifts in pulse
profile and spin-down luminosity \citep{Lyne10}, and dramatic, rapid
shifts in magnetospheric configuration \citep{Hermsen13}.  More recently,
\citet{Allafort13} presented the discovery of the first state switch
observed in \g-rays, in which an abrupt increase of the spin-down rate
\nudot{} of PSR~J2021$+$4026 was accompanied by both a decrease in
total flux and change in pulse shape.

The efficacy of pulsar timing is reduced by {timing noise} (TN), a red
noise process first recognized by \citet{Boynton72}.  TN is intrinsic
to pulsar rotation and operates over a broad range of time-scales.  It
is generally strongest in young pulsars \citep[e.g.][]{Shannon10},
with noise processes whose amplitudes can be far larger than the
pulsar period ($>$1\,s), while TN in MSPs is much weaker, with
amplitudes of order 1\,$\upmu$s.  Because TN cannot be modelled
\textit{a priori}, it confuses the deterministic signature of other
parameters of interest.  On the other hand, if TN can be definitely
related to a particular physical mechanism
\citep[e.g.][]{Cordes81,Lyne10}, it provides another means of
understanding neutron stars and their magnetospheres.

For \fm{} pulsars, TN is the primary obstacle to identifying
counterparts at other wavelengths.  Because the angular resolution of
the LAT is coarse compared to focusing telescopes, determining the
counterpart of a \g-ray source typically requires the detection of
correlated temporal variability \citep{Nolan12}.  On the other hand,
pulsar timing offers an essentially interferometric localization with
a baseline of up to 1\,AU, with position errors inducing an annual
sinusoidal modulation in the pulse TOAs.  TN obscures and corrupts
this signal, but a data set spanning years has multiple realizations
of both the sinusoid and the annual-scale fluctuations of the noise
process, allowing the signals to be disentangled. Thus, it behooves us
to collect long sets of precise TOAs for each detected \g-ray pulsar.

As with modelling TN, measuring TOAs with \fm{} also presents
challenges.  The LAT only collects a few photons per day from a
typical \g-ray pulsar, and measurement of a single TOA requires
integrations ranging from hours to months.  Sources are embedded in
the strong diffuse background and are sometimes confused with bright
neighboring sources; both situations require sophisticated spectral
modelling to optimize the pulsar signal.  Finally, the LAT pulsar
sample includes many young, energetic pulsars with strong TN and
glitches.  Glitches are sudden step changes in $\nu$ and \nudot{},
often with a transient component in which the increased spin-down rate
``recovers'' nearly to the pre-glitch value; see e.g.
\citet{Espinoza11} for a large sample of examples.  All told, during
the time required to obtain a single TOA, the phase of the neutron
star may have drifted considerably, and one finds oneself in the
unenviable position of needing the timing model to measure the timing
model!

With the large pulsar population and the steady acquisition of more
than five years of all-sky data, we are now in a position to better
address both TN and TOA measurements.  Building on the work of
\citet{Ray11}, we have developed a suite of techniques for reliably
measuring TOAs from \fm{} data with maximum likelihood.  By applying a
relatively new approach, in which TN is modelled as correlations among
TOAs \citep[][hereafter C11]{Coles11}, we are able to accurately
measure the TN component and obtain unbiased estimates for pulsar
timing parameters.  We have incorporated these techniques into a
pipeline which provides updated timing solutions and TOAs for all
\g-ray pulsars above a nominal flux threshold.  These
products---including, for the first time, TOAs---are freely available
to the community.  We expect them to be of interest to researchers for
a variety of topics, including the construction of detailed folded
light curves, population analyses, and searches for state changes.

We introduce the timing sample and the reduction of \g-ray and radio
data in \S\ref{sec:sample}, \ref{sec:lat_reduction}, and
\ref{sec:radio_reduction}, respectively.  We describe the maximum
likelihood-based timing pipeline and resulting data products in
\S\ref{sec:pulsar_timing} and \S\ref{sec:pipeline}.  In
\S\ref{sec:ml}, we discuss in detail our approach to measuring TN.
With those results, we present an analysis of TN in the \g-ray pulsar
population (\S\ref{sec:model_selection} and \S\ref{sec:measurements}),
in particular characterizing the strength and shape of the TN
spectrum.  Finally, we demonstrate the power of our method by
determining reliable, precise timing positions for the sample in
\S\ref{sec:astrometry}.  Where appropriate, we discuss results
\textit{in situ}.

\section{Analysis}
\label{sec:analysis}

\subsection{Pulsar Sample}
\label{sec:sample}

Our approximately flux-limited sample comprises \nump{} pulsars,
including 36 radio-quiet or radio-faint pulsars which are too dim to
time with modern radio telescopes, viz. PSRs~J1124$-$5916 and
J1907$+$0602.  This sample represents a marked increase over the 16
pulsars characterized in \citet{Ray11}, with gains coming from both a
larger pool of detected \g-ray pulsars and a flux threshold lower by a
factor of about five. 

The basic criterion for selection is that the pulse is reliably
detected in the TOA integration interval.  We selected as our maximum
interval a span of 56 days which is approximately the precessional
period of the \fm{} spacecraft, on which the exposure to a source
cycles.  Because sharp pulses are more easily detected than broad
ones, this detectability criterion does not translate directly into a
flux cutoff.  Empirically, if a pulsar can generate a weighted
$H$-test statistic \citep{Kerr11,deJager89} of $\sim$15, about
3$\sigma$, during a given interval, it can be timed reliably at that
cadence.  We based our thresholds on the 2PC weighted $H$-test values,
computed with three years of data: for a cadence of 56 days, pulsars
reaching $H$$\sim$250 in three years satisfy the detection criterion.
We were unable to obtain phase-connected timing solutions for a few
pulsars surpassing this threshold, viz.  J1838$-$0537, J0554$+$3107,
J1810$+$1744, and J2043$+$2740, and these pulsars are excluded from
the sample.  These timing solutions may be recovered with unbinned
timing techniques, improved \fm{} event reconstruction, and/or
contributions of radio TOAs.

We conclude with a few remarks on our choice of maximum cadence.  As
we describe below, extraction of a TOA requires a timing solution
valid over the course of the integration period.  In the case of young
pulsars, both TN and glitches are prohibitive of long integration
periods, with our experience leading us to adopt 56 days.  On the
other hand, MSPs, with their much more stable rotation, can benefit
from even longer integration periods, effectively lowering the sample
flux threshold.  However, these longer integration periods prohibit
timing of short time-scale features, e.g. the orbital period,
requiring ancillary observations from radio telescopes to provide
initial characterization.  Ultimately, these issues demand an unbinned
analysis, in which the timing model is fitted directly to photon
timestamps.  Because pulsar timing machinery is based on the concept
of TOA, such a switch represents a major task for future workers.

\subsection{LAT Data Reduction}
\label{sec:lat_reduction}

We analyze 5.2 years of Source-class Pass 7 LAT data acquired between
2008 Aug 4 (MET 239557418) and 2013 Oct 21 (MET 404063018).  We note
that these data are ``unreprocessed'' \citep[cf.][]{Johnson13}, but
that the improved properties of the reprocessed data have little
benefit for pulsar timing.  The data span nearly the entire period of
the uniform \fm{} all-sky survey, terminating only a month before the
switch into a modified survey mode with enhanced exposure to the
Galactic center and modestly compromised exposure elsewhere.

Using \textit{gtmktime} (\fm{} Science Tools version 09-31-02), we
filtered events lying outside the Good Time Intervals defined by the
\fm{}-LAT team, which exclude the regular passages through the South
Atlantic Anomaly and transient events such as bright solar flares.  We
also reject events when the rocking angle of the spacecraft exceeds
$52^{\circ}$ and those with a zenith angle $>$100$^{\circ}$; these
cuts primarily eliminate background from \g-rays generated along the
earth's limb.  To further reduce the background, primarily from the
bright Galactic diffuse emission, we apply photon weighting
\citep{Kerr11} with {\it gtsrcprob} using the \texttt{P7SOURCE\_V6}
instrument response function.  The photon weights (see
\S\ref{sec:pulsar_timing} for application) are estimates of the
probability that a given event arrives from the pulsar instead of the
background, and require an estimate for the spectrum of each.  For
most pulsars, we use the 2PC spectral and background
models\footnote{http://fermi.gsfc.nasa.gov/ssc/data/access/lat/BackgroundModels.html}.
For four additional pulsars discovered in blind searches,
PSRs~J0554$+$3107, J1422$-$6138, J1522$-$5735, J1932$+$1916, we use
the spectral models derived in \citet{Pletsch13}.  Although spectral
models may be derived with different data sets and instrumental
representations than those employed here, the photon weights needed
for timing analysis do not depend sensitively on the precision of the
model \citep{Kerr11}.

Finally, we select photons within 2$\arcdeg$ of the pulsar position
with reconstructed energies lying between 100\,MeV and 30\,GeV.  Where
required, we make use of the \textit{Fermi} plugin for \textsc{Tempo2}
\citep{Hobbs06} to assign pulse phase to these photons.

\subsection{Radio Data Reduction}
\label{sec:radio_reduction}


For a subset of our pulsars, we have obtained high-quality timing data
from radio telescopes, primarily from ongoing programs to monitor
high-$\dot{E}$ pulsars in support the \fm{} mission \citep{Smith08}.
Most data come from the Parkes campaign \citep{Weltevrede10} and were
reduced via a pipeline that performs polarimetric calibration and
impulsive and narrowband RFI excision; the raw data are available
through the CSIRO Data Archive Portal\footnote{https://data.csiro.au}
\citep{Hobbs11}.

Two MSPs (J1902$-$5105 and J1514$-$4946) discovered in a Parkes survey
were monitored for two years and we include the TOAs, produced and
reduced as per \citet{Kerr12}, in our sample.

Two young pulsars, J1747$-$2958 and J1833$-$1034, are visible to
Parkes but too dim to be reliably timed.  These were monitored with
the Green Bank Telescope for three years, and we likewise include
these data.

\begin{deluxetable*}{crcllll}
\tablewidth{\textwidth}
\tablecaption{\label{tab:overview_normal}Positions from \fm{} timing
and from multi-wavelength (MWL) observations for normal (`unrecycled')
pulsars.  All positions are computed at the epoch MJD 55555 (25 Dec
2010), and positions reported in the literature are advanced to this
epoch, with propagation of error, when proper motions are available.
The TOA cadence and best-fit timing noise models are also shown.\\Astrometry references: $^\mathrm{a}$\citet{Halpern04}; $^\mathrm{b}$\citet{Slane02}; $^\mathrm{c}$\citet{DeLuca11}; $^\mathrm{d}$\citet{Kaplan08}; $^\mathrm{e}$\citet{Ray11}; $^\mathrm{f}$\citet{Caraveo98}; $^\mathrm{g}$\citet{Faherty07}; $^\mathrm{h}$\citet{Brisken03}; $^\mathrm{i}$\citet{Dodson03}; $^\mathrm{j}$\citet{Kargaltsev12}; $^\mathrm{k}$\citet{Keith08}; $^\mathrm{l}$\citet{Stappers99}; $^\mathrm{m}$\citet{Gonzalez06}; $^\mathrm{n}$\citet{Mignani10}; $^\mathrm{o}$\citet{Hughes03}; $^\mathrm{p}$\citet{Marelli12}; $^\mathrm{q}$\citet{Camilo04}; $^\mathrm{r}$\citet{Ng05}; $^\mathrm{s}$\citet{Kargaltsev08}; $^\mathrm{j}$\citet{Kargaltsev12}; $^\mathrm{e}$\citet{Ray11}; $^\mathrm{t}$\citet{Romani10}; $^\mathrm{u}$\citet{Gaensler04}; $^\mathrm{v}$\citet{VanEtten12}; $^\mathrm{e}$\citet{Ray11}; $^\mathrm{w}$\citet{Camilo06}; $^\mathrm{x}$\citet{Halpern02}; $^\mathrm{y}$\citet{Abdo10_1907}; $^\mathrm{A}$\citet{Zeiger08}; $^\mathrm{j}$\citet{Kargaltsev12}; $^\mathrm{B}$\citet{VanEtten08}; $^\mathrm{C}$\citet{Weisskopf11}; $^\mathrm{D}$\citet{Camilo09}; $^\mathrm{E}$\citet{Halpern01}}
\tablecolumns{7}
\tablehead{
\colhead{Name} &
\colhead{Cadence} &
\colhead{TN Model} &
\colhead{R.A.} &
\colhead{Decl.} &
\colhead{MWL R.A.} &
\colhead{MWL Decl.} \\
\colhead{} &
\colhead{(days)} &
\colhead{} &
\colhead{(J2000.0)} &
\colhead{(J2000.0)} &
\colhead{(J2000.0)} &
\colhead{(J2000.0)}
}
\startdata
J0007$+$7303 & 7  & BPL & 00$^{\mathrm{h}}$07$^{\mathrm{m}}$00$\fs$205 (2.382) & $+$73$\degr$03$\arcmin$17$\farcs$46 (9.93) & 00$^{\mathrm{h}}$07$^{\mathrm{m}}$01$\fs$560 (0.010) & $+$73$\degr$03$\arcmin$08$\farcs$10 (0.10)$^\mathrm{a}$\\
J0106$+$4855 & 28  & BPL & 01$^{\mathrm{h}}$06$^{\mathrm{m}}$25$\fs$030 (0.046) & $+$48$\degr$55$\arcmin$52$\farcs$01 (0.57) & -- & --\\
J0205$+$6449 & 14  & PL & 02$^{\mathrm{h}}$05$^{\mathrm{m}}$35$\fs$985 (2.162) & $+$64$\degr$49$\arcmin$32$\farcs$95 (15.76) & 02$^{\mathrm{h}}$05$^{\mathrm{m}}$37$\fs$920 (0.020) & $+$64$\degr$49$\arcmin$42$\farcs$80 (0.72)$^\mathrm{b}$\\
J0248$+$6021 & 56  & BPL & 02$^{\mathrm{h}}$48$^{\mathrm{m}}$18$\fs$642 (0.160) & $+$60$\degr$21$\arcmin$34$\farcs$70 (1.50) & -- & --\\
J0357$+$3205 & 14  & BPL & 03$^{\mathrm{h}}$57$^{\mathrm{m}}$52$\fs$161 (0.075) & $+$32$\degr$05$\arcmin$23$\farcs$63 (3.37) & 03$^{\mathrm{h}}$57$^{\mathrm{m}}$52$\fs$329 (0.017) & $+$32$\degr$05$\arcmin$20$\farcs$73 (0.25)$^\mathrm{c}$\\
J0534$+$2200 & 7  & BPL & 05$^{\mathrm{h}}$34$^{\mathrm{m}}$31$\fs$889 (0.088) & $+$22$\degr$01$\arcmin$18$\farcs$46 (24.76) & 05$^{\mathrm{h}}$34$^{\mathrm{m}}$31$\fs$929 (0.004) & $+$22$\degr$00$\arcmin$52$\farcs$16 (0.06)$^\mathrm{d}$\\
J0622$+$3749 & 56  & FPL & 06$^{\mathrm{h}}$22$^{\mathrm{m}}$10$\fs$411 (0.073) & $+$37$\degr$49$\arcmin$14$\farcs$60 (3.49) & -- & --\\
J0631$+$1036 & 56  & BPL & 06$^{\mathrm{h}}$31$^{\mathrm{m}}$27$\fs$369 (0.113) & $+$10$\degr$37$\arcmin$05$\farcs$63 (7.03) & -- & --\\
J0633$+$0632 & 7  & BPL & 06$^{\mathrm{h}}$33$^{\mathrm{m}}$44$\fs$136 (0.025) & $+$06$\degr$32$\arcmin$29$\farcs$48 (1.30) & 06$^{\mathrm{h}}$33$^{\mathrm{m}}$44$\fs$142 (0.020) & $+$06$\degr$32$\arcmin$30$\farcs$40 (0.30)$^\mathrm{e}$\\
J0633$+$1746 & 14  & BPL & 06$^{\mathrm{h}}$33$^{\mathrm{m}}$54$\fs$288 (0.003) & $+$17$\degr$46$\arcmin$15$\farcs$32 (0.44) & 06$^{\mathrm{h}}$33$^{\mathrm{m}}$54$\fs$310 (0.003) & $+$17$\degr$46$\arcmin$14$\farcs$60 (0.04)$^\mathrm{f}$$^\mathrm{g}$\\
J0659$+$1414 & 28  & BPL & 06$^{\mathrm{h}}$59$^{\mathrm{m}}$48$\fs$183 (0.005) & $+$14$\degr$14$\arcmin$21$\farcs$54 (0.41) & 06$^{\mathrm{h}}$59$^{\mathrm{m}}$48$\fs$180 (0.001) & $+$14$\degr$14$\arcmin$21$\farcs$13 (0.00)$^\mathrm{h}$\\
J0734$-$1559 & 56  & BPL & 07$^{\mathrm{h}}$34$^{\mathrm{m}}$45$\fs$693 (0.028) & $-$15$\degr$59$\arcmin$18$\farcs$43 (0.67) & -- & --\\
J0835$-$4510 & 7  & BPL & 08$^{\mathrm{h}}$35$^{\mathrm{m}}$20$\fs$612
(0.116) & $-$45$\degr$10$\arcmin$33$\farcs$98 (1.35) &
08$^{\mathrm{h}}$35$^{\mathrm{m}}$20$\fs$560 ($<$0.001) &
$-$45$\degr$10$\arcmin$34$\farcs$55 ($<$0.01)$^\mathrm{i}$\\
J0908$-$4913 & 56  & BPL & 09$^{\mathrm{h}}$08$^{\mathrm{m}}$35$\fs$420 (0.023) & $-$49$\degr$13$\arcmin$05$\farcs$79 (0.24) & 09$^{\mathrm{h}}$08$^{\mathrm{m}}$35$\fs$393 (0.020) & $-$49$\degr$13$\arcmin$04$\farcs$84 (0.30)$^\mathrm{j}$\\
J1016$-$5857 & 56  & PL & 10$^{\mathrm{h}}$16$^{\mathrm{m}}$21$\fs$266 (0.092) & $-$58$\degr$57$\arcmin$11$\farcs$70 (0.71) & -- & --\\
J1023$-$5746 & 7  & BPL & 10$^{\mathrm{h}}$23$^{\mathrm{m}}$02$\fs$388 (0.703) & $-$57$\degr$46$\arcmin$09$\farcs$86 (5.39) & -- & --\\
J1028$-$5819 & 7  & BPL & 10$^{\mathrm{h}}$28$^{\mathrm{m}}$27$\fs$888 (0.004) & $-$58$\degr$19$\arcmin$06$\farcs$14 (0.03) & 10$^{\mathrm{h}}$28$^{\mathrm{m}}$28$\fs$000 (0.100) & $-$58$\degr$19$\arcmin$05$\farcs$20 (1.50)$^\mathrm{k}$\\
J1044$-$5737 & 14  & BPL & 10$^{\mathrm{h}}$44$^{\mathrm{m}}$32$\fs$791 (0.024) & $-$57$\degr$37$\arcmin$19$\farcs$39 (0.17) & -- & --\\
J1048$-$5832 & 7  & BPL & 10$^{\mathrm{h}}$48$^{\mathrm{m}}$12$\fs$391 (0.253) & $-$58$\degr$32$\arcmin$02$\farcs$68 (1.85) & 10$^{\mathrm{h}}$48$^{\mathrm{m}}$12$\fs$640 (0.040) & $-$58$\degr$32$\arcmin$03$\farcs$75 (0.01)$^\mathrm{l}$$^\mathrm{m}$\\
J1057$-$5226 & 14  & BPL & 10$^{\mathrm{h}}$57$^{\mathrm{m}}$58$\fs$970 (0.007) & $-$52$\degr$26$\arcmin$56$\farcs$52 (0.07) & 10$^{\mathrm{h}}$57$^{\mathrm{m}}$58$\fs$978 (0.010) & $-$52$\degr$26$\arcmin$56$\farcs$27 (0.15)$^\mathrm{n}$\\
J1124$-$5916 & 14  & PL & 11$^{\mathrm{h}}$24$^{\mathrm{m}}$39$\fs$174 (0.863) & $-$59$\degr$16$\arcmin$15$\farcs$32 (6.83) & 11$^{\mathrm{h}}$24$^{\mathrm{m}}$39$\fs$100 (0.100) & $-$59$\degr$16$\arcmin$20$\farcs$00 (1.00)$^\mathrm{o}$\\
J1135$-$6055 & 28  & BPL & 11$^{\mathrm{h}}$35$^{\mathrm{m}}$08$\fs$428 (0.048) & $-$60$\degr$55$\arcmin$36$\farcs$50 (0.35) & 11$^{\mathrm{h}}$35$^{\mathrm{m}}$08$\fs$450 (0.100) & $-$60$\degr$55$\arcmin$36$\farcs$99 (1.70)$^\mathrm{p}$\\
J1357$-$6429 & 56  & BPL & 13$^{\mathrm{h}}$57$^{\mathrm{m}}$02$\fs$349 (0.278) & $-$64$\degr$29$\arcmin$31$\farcs$67 (2.09) & 13$^{\mathrm{h}}$57$^{\mathrm{m}}$02$\fs$430 (0.020) & $-$64$\degr$29$\arcmin$30$\farcs$20 (0.10)$^\mathrm{q}$\\
J1413$-$6205 & 14  & BPL & 14$^{\mathrm{h}}$13$^{\mathrm{m}}$30$\fs$135 (0.113) & $-$62$\degr$05$\arcmin$37$\farcs$02 (0.94) & -- & --\\
J1418$-$6058 & 7  & PL & 14$^{\mathrm{h}}$18$^{\mathrm{m}}$42$\fs$725 (0.134) & $-$60$\degr$58$\arcmin$01$\farcs$89 (1.27) & 14$^{\mathrm{h}}$18$^{\mathrm{m}}$42$\fs$700 (0.080) & $-$60$\degr$58$\arcmin$03$\farcs$10 (0.40)$^\mathrm{r}$\\
J1420$-$6048 & 28  & PL & 14$^{\mathrm{h}}$20$^{\mathrm{m}}$08$\fs$163 (0.250) & $-$60$\degr$48$\arcmin$15$\farcs$55 (2.32) & -- & --\\
J1429$-$5911 & 28  & BPL & 14$^{\mathrm{h}}$29$^{\mathrm{m}}$58$\fs$564 (0.019) & $-$59$\degr$11$\arcmin$36$\farcs$02 (0.18) & -- & --\\
J1459$-$6053 & 14  & BPL & 14$^{\mathrm{h}}$59$^{\mathrm{m}}$30$\fs$222 (0.024) & $-$60$\degr$53$\arcmin$19$\farcs$87 (0.25) & -- & --\\
J1509$-$5850 & 28  & BPL & 15$^{\mathrm{h}}$09$^{\mathrm{m}}$27$\fs$153 (0.004) & $-$58$\degr$50$\arcmin$56$\farcs$05 (0.05) & 15$^{\mathrm{h}}$09$^{\mathrm{m}}$27$\fs$163 (0.020) & $-$58$\degr$50$\arcmin$56$\farcs$12 (0.30)$^\mathrm{s}$\\
J1522$-$5735 & 56  & BPL & 15$^{\mathrm{h}}$22$^{\mathrm{m}}$05$\fs$226 (0.107) & $-$57$\degr$35$\arcmin$00$\farcs$91 (1.19) & -- & --\\
J1620$-$4927 & 28  & BPL & 16$^{\mathrm{h}}$20$^{\mathrm{m}}$41$\fs$594 (0.089) & $-$49$\degr$27$\arcmin$36$\farcs$21 (1.70) & -- & --\\
J1709$-$4429 & 7  & BPL & 17$^{\mathrm{h}}$09$^{\mathrm{m}}$42$\fs$834 (0.339) & $-$44$\degr$28$\arcmin$59$\farcs$26 (9.41) & 17$^{\mathrm{h}}$09$^{\mathrm{m}}$42$\fs$748 (0.000) & $-$44$\degr$29$\arcmin$08$\farcs$37 (0.00)\\
J1718$-$3825 & 28  & BPL & 17$^{\mathrm{h}}$18$^{\mathrm{m}}$13$\fs$576 (0.024) & $-$38$\degr$25$\arcmin$18$\farcs$56 (1.07) & 17$^{\mathrm{h}}$18$^{\mathrm{m}}$13$\fs$541 (0.020) & $-$38$\degr$25$\arcmin$17$\farcs$26 (0.30)$^\mathrm{j}$\\
J1732$-$3131 & 14  & BPL & 17$^{\mathrm{h}}$32$^{\mathrm{m}}$33$\fs$551 (0.031) & $-$31$\degr$31$\arcmin$20$\farcs$85 (2.59) & 17$^{\mathrm{h}}$32$^{\mathrm{m}}$33$\fs$551 (0.030) & $-$31$\degr$31$\arcmin$23$\farcs$92 (0.50)$^\mathrm{e}$\\
J1741$-$2054 & 14  & BPL & 17$^{\mathrm{h}}$41$^{\mathrm{m}}$57$\fs$282 (0.026) & $-$20$\degr$54$\arcmin$04$\farcs$11 (6.71) & 17$^{\mathrm{h}}$41$^{\mathrm{m}}$57$\fs$280 (0.020) & $-$20$\degr$54$\arcmin$11$\farcs$80 (0.30)$^\mathrm{t}$\\
J1746$-$3239 & 56  & PL & 17$^{\mathrm{h}}$46$^{\mathrm{m}}$55$\fs$143 (0.141) & $-$32$\degr$39$\arcmin$36$\farcs$36 (9.67) & -- & --\\
J1747$-$2958 & 28  & BPL & 17$^{\mathrm{h}}$47$^{\mathrm{m}}$15$\fs$869 (0.018) & $-$29$\degr$58$\arcmin$00$\farcs$93 (1.97) & 17$^{\mathrm{h}}$47$^{\mathrm{m}}$15$\fs$854 (0.004) & $-$29$\degr$58$\arcmin$01$\farcs$38 (0.04)$^\mathrm{u}$\\
J1803$-$2149 & 56  & BPL & 18$^{\mathrm{h}}$03$^{\mathrm{m}}$09$\fs$632 (0.007) & $-$21$\degr$49$\arcmin$00$\farcs$87 (3.66) & -- & --\\
J1809$-$2332 & 14  & BPL & 18$^{\mathrm{h}}$09$^{\mathrm{m}}$50$\fs$227 (0.018) & $-$23$\degr$32$\arcmin$12$\farcs$98 (14.41) & 18$^{\mathrm{h}}$09$^{\mathrm{m}}$50$\fs$249 (0.030) & $-$23$\degr$32$\arcmin$22$\farcs$67 (0.10)$^\mathrm{v}$\\
J1813$-$1246 & 14  & BPL & 18$^{\mathrm{h}}$13$^{\mathrm{m}}$23$\fs$768 (0.005) & $-$12$\degr$46$\arcmin$00$\farcs$63 (0.40) & -- & --\\
J1826$-$1256 & 7  & BPL & 18$^{\mathrm{h}}$26$^{\mathrm{m}}$08$\fs$570 (0.042) & $-$12$\degr$56$\arcmin$33$\farcs$53 (3.05) & 18$^{\mathrm{h}}$26$^{\mathrm{m}}$08$\fs$540 (0.010) & $-$12$\degr$56$\arcmin$34$\farcs$60 (0.10)$^\mathrm{e}$\\
J1833$-$1034 & 28  & BPL & 18$^{\mathrm{h}}$33$^{\mathrm{m}}$33$\fs$612 (0.047) & $-$10$\degr$34$\arcmin$07$\farcs$69 (3.20) & 18$^{\mathrm{h}}$33$^{\mathrm{m}}$33$\fs$570 (0.020) & $-$10$\degr$34$\arcmin$07$\farcs$50 (0.10)$^\mathrm{w}$\\
J1836$+$5925 & 14  & FPL & 18$^{\mathrm{h}}$36$^{\mathrm{m}}$13$\fs$661 (0.013) & $+$59$\degr$25$\arcmin$29$\farcs$77 (0.11) & 18$^{\mathrm{h}}$36$^{\mathrm{m}}$13$\fs$723 (0.017) & $+$59$\degr$25$\arcmin$30$\farcs$05 (0.10)$^\mathrm{x}$\\
J1846$+$0919 & 56  & FPL & 18$^{\mathrm{h}}$46$^{\mathrm{m}}$25$\fs$864 (0.038) & $+$09$\degr$19$\arcmin$48$\farcs$99 (1.08) & -- & --\\
J1907$+$0602 & 7  & PL & 19$^{\mathrm{h}}$07$^{\mathrm{m}}$54$\fs$751 (0.044) & $+$06$\degr$02$\arcmin$15$\farcs$79 (1.32) & 19$^{\mathrm{h}}$07$^{\mathrm{m}}$54$\fs$760 (0.050) & $+$06$\degr$02$\arcmin$14$\farcs$60 (0.70)$^\mathrm{y}$\\
J1932$+$1916 & 56  & BPL & 19$^{\mathrm{h}}$32$^{\mathrm{m}}$19$\fs$775 (0.063) & $+$19$\degr$16$\arcmin$38$\farcs$06 (1.56) & -- & --\\
J1952$+$3252 & 7  & PL & 19$^{\mathrm{h}}$52$^{\mathrm{m}}$58$\fs$181 (0.046) & $+$32$\degr$52$\arcmin$40$\farcs$98 (0.72) & 19$^{\mathrm{h}}$52$^{\mathrm{m}}$58$\fs$189 (0.001) & $+$32$\degr$52$\arcmin$40$\farcs$40 (0.01)$^\mathrm{A}$\\
J1954$+$2836 & 28  & BPL & 19$^{\mathrm{h}}$54$^{\mathrm{m}}$19$\fs$144 (0.008) & $+$28$\degr$36$\arcmin$04$\farcs$84 (0.12) & -- & --\\
J1957$+$5033 & 28  & FPL & 19$^{\mathrm{h}}$57$^{\mathrm{m}}$38$\fs$391 (0.068) & $+$50$\degr$33$\arcmin$21$\farcs$24 (0.69) & -- & --\\
J1958$+$2846 & 14  & BPL & 19$^{\mathrm{h}}$58$^{\mathrm{m}}$40$\fs$064 (0.013) & $+$28$\degr$45$\arcmin$54$\farcs$07 (0.22) & 19$^{\mathrm{h}}$58$^{\mathrm{m}}$40$\fs$013 (0.020) & $+$28$\degr$45$\arcmin$55$\farcs$12 (0.30)$^\mathrm{j}$\\
J2021$+$3651 & 7  & BPL & 20$^{\mathrm{h}}$21$^{\mathrm{m}}$05$\fs$501 (0.150) & $+$36$\degr$51$\arcmin$03$\farcs$79 (2.15) & 20$^{\mathrm{h}}$21$^{\mathrm{m}}$05$\fs$430 (0.030) & $+$36$\degr$51$\arcmin$04$\farcs$63 (0.50)$^\mathrm{B}$\\
J2021$+$4026 & 14  & BPL & 20$^{\mathrm{h}}$21$^{\mathrm{m}}$30$\fs$496 (0.502) & $+$40$\degr$26$\arcmin$53$\farcs$56 (6.16) & 20$^{\mathrm{h}}$21$^{\mathrm{m}}$30$\fs$733 (0.020) & $+$40$\degr$26$\arcmin$46$\farcs$04 (0.30)$^\mathrm{C}$\\
J2028$+$3332 & 28  & FPL & 20$^{\mathrm{h}}$28$^{\mathrm{m}}$19$\fs$876 (0.008) & $+$33$\degr$32$\arcmin$04$\farcs$23 (0.12) & -- & --\\
J2030$+$3641 & 56  & FPL & 20$^{\mathrm{h}}$30$^{\mathrm{m}}$00$\fs$241 (0.022) & $+$36$\degr$41$\arcmin$27$\farcs$30 (0.29) & -- & --\\
J2030$+$4415 & 56  & BPL & 20$^{\mathrm{h}}$30$^{\mathrm{m}}$51$\fs$396 (0.021) & $+$44$\degr$15$\arcmin$38$\farcs$73 (0.26) & -- & --\\
J2032$+$4127 & 7  & BPL & 20$^{\mathrm{h}}$32$^{\mathrm{m}}$12$\fs$932 (0.344) & $+$41$\degr$27$\arcmin$22$\farcs$82 (4.47) & 20$^{\mathrm{h}}$32$^{\mathrm{m}}$13$\fs$143 (0.024) & $+$41$\degr$27$\arcmin$24$\farcs$54 (0.27)$^\mathrm{D}$\\
J2055$+$2539 & 28  & FPL & 20$^{\mathrm{h}}$55$^{\mathrm{m}}$48$\fs$948 (0.032) & $+$25$\degr$39$\arcmin$58$\farcs$93 (0.62) & -- & --\\
J2111$+$4606 & 28  & BPL & 21$^{\mathrm{h}}$11$^{\mathrm{m}}$24$\fs$125 (0.069) & $+$46$\degr$06$\arcmin$30$\farcs$66 (0.71) & -- & --\\
J2139$+$4716 & 56  & BPL & 21$^{\mathrm{h}}$39$^{\mathrm{m}}$55$\fs$878 (0.064) & $+$47$\degr$16$\arcmin$13$\farcs$12 (0.68) & -- & --\\
J2229$+$6114 & 7  & BPL & 22$^{\mathrm{h}}$29$^{\mathrm{m}}$05$\fs$262 (0.515) & $+$61$\degr$14$\arcmin$08$\farcs$48 (3.96) & 22$^{\mathrm{h}}$29$^{\mathrm{m}}$05$\fs$280 (0.070) & $+$61$\degr$14$\arcmin$09$\farcs$30 (0.50)$^\mathrm{E}$\\
J2238$+$5903 & 14  & BPL & 22$^{\mathrm{h}}$38$^{\mathrm{m}}$28$\fs$089 (0.187) & $+$59$\degr$03$\arcmin$43$\farcs$37 (1.46) & -- & --\\
\enddata
\end{deluxetable*}

\begin{deluxetable*}{crcllll}
\tablewidth{\textwidth}
\tablecaption{\label{tab:overview_msp}As Table \ref{tab:overview_normal} for millisecond (`recycled') pulsars.  \\Astrometry references: $^\mathrm{a}$\citet{Deller08}}
\tablecolumns{7}
\tablehead{
\colhead{Name} &
\colhead{Cadence} &
\colhead{TN Model} &
\colhead{R.A.} &
\colhead{Decl.} &
\colhead{MWL R.A.} &
\colhead{MWL Decl.} \\
\colhead{} &
\colhead{(days)} &
\colhead{} &
\colhead{(J2000.0)} &
\colhead{(J2000.0)} &
\colhead{(J2000.0)} &
\colhead{(J2000.0)}
}
\startdata
J0030$+$0451 & 28  & FPL & 00$^{\mathrm{h}}$30$^{\mathrm{m}}$27$\fs$4277 (0.0010) & $+$04$\degr$51$\arcmin$39$\farcs$710 (0.035) & -- & --\\
J0218$+$4232 & 56  & FPL & 02$^{\mathrm{h}}$18$^{\mathrm{m}}$06$\fs$3604 (0.0008) & $+$42$\degr$32$\arcmin$17$\farcs$365 (0.015) & -- & --\\
J0340$+$4130 & 56  & BPL & 03$^{\mathrm{h}}$40$^{\mathrm{m}}$23$\fs$2887 (0.0004) & $+$41$\degr$30$\arcmin$45$\farcs$300 (0.012) & -- & --\\
J0437$-$4715 & 28  & FPL &
04$^{\mathrm{h}}$37$^{\mathrm{m}}$15$\fs$9314 (0.0007) &
$-$47$\degr$15$\arcmin$09$\farcs$318 (0.008) &
04$^{\mathrm{h}}$37$^{\mathrm{m}}$15$\fs$9309 ($<$0.0001) &
$-$47$\degr$15$\arcmin$09$\farcs$318 ($<$0.001)$^\mathrm{a}$\\
J0613$-$0200 & 28  & FPL & 06$^{\mathrm{h}}$13$^{\mathrm{m}}$43$\fs$9759 (0.0002) & $-$02$\degr$00$\arcmin$47$\farcs$228 (0.010) & -- & --\\
J0614$-$3329 & 28  & NTN & 06$^{\mathrm{h}}$14$^{\mathrm{m}}$10$\fs$3479 (0.0000) & $-$33$\degr$29$\arcmin$54$\farcs$116 (0.001) & -- & --\\
J0751$+$1807 & 56  & FPL & 07$^{\mathrm{h}}$51$^{\mathrm{m}}$09$\fs$1539 (0.0020) & $+$18$\degr$07$\arcmin$38$\farcs$335 (0.154) & -- & --\\
J1231$-$1411 & 28  & NTN & 12$^{\mathrm{h}}$31$^{\mathrm{m}}$11$\fs$3069 (0.0001) & $-$14$\degr$11$\arcmin$43$\farcs$628 (0.003) & -- & --\\
J1514$-$4946 & 58  & FPL & 15$^{\mathrm{h}}$14$^{\mathrm{m}}$19$\fs$1142 (0.0001) & $-$49$\degr$46$\arcmin$15$\farcs$528 (0.003) & -- & --\\
J1614$-$2230 & 28  & FPL & 16$^{\mathrm{h}}$14$^{\mathrm{m}}$36$\fs$5070 (0.0016) & $-$22$\degr$30$\arcmin$31$\farcs$207 (0.120) & -- & --\\
J1744$-$1134 & 56  & FPL & 17$^{\mathrm{h}}$44$^{\mathrm{m}}$29$\fs$4103 (0.0004) & $-$11$\degr$34$\arcmin$54$\farcs$727 (0.031) & -- & --\\
J1902$-$5105 & 56  & FPL & 19$^{\mathrm{h}}$02$^{\mathrm{m}}$02$\fs$8476 (0.0001) & $-$51$\degr$05$\arcmin$56$\farcs$969 (0.002) & -- & --\\
J1959$+$2048 & 56  & FPL & 19$^{\mathrm{h}}$59$^{\mathrm{m}}$36$\fs$7480 (0.0001) & $+$20$\degr$48$\arcmin$14$\farcs$599 (0.002) & -- & --\\
J2017$+$0603 & 28  & FPL & 20$^{\mathrm{h}}$17$^{\mathrm{m}}$22$\fs$7048 (0.0003) & $+$06$\degr$03$\arcmin$05$\farcs$585 (0.008) & -- & --\\
J2043$+$1711 & 56  & FPL & 20$^{\mathrm{h}}$43$^{\mathrm{m}}$20$\fs$8828 (0.0002) & $+$17$\degr$11$\arcmin$28$\farcs$941 (0.005) & -- & --\\
J2124$-$3358 & 28  & FPL & 21$^{\mathrm{h}}$24$^{\mathrm{m}}$43$\fs$8464 (0.0010) & $-$33$\degr$58$\arcmin$44$\farcs$961 (0.028) & -- & --\\
J2214$+$3000 & 56  & FPL & 22$^{\mathrm{h}}$14$^{\mathrm{m}}$38$\fs$8480 (0.0004) & $+$30$\degr$00$\arcmin$38$\farcs$209 (0.007) & -- & --\\
J2241$-$5236 & 56  & FPL & 22$^{\mathrm{h}}$41$^{\mathrm{m}}$42$\fs$0199 (0.0002) & $-$52$\degr$36$\arcmin$36$\farcs$224 (0.003) & -- & --\\
J2302$+$4442 & 56  & NTN & 23$^{\mathrm{h}}$02$^{\mathrm{m}}$46$\fs$9789 (0.0002) & $+$44$\degr$42$\arcmin$22$\farcs$102 (0.003) & -- & --\\
\enddata
\end{deluxetable*}

\subsection{Pulsar Timing}
\label{sec:pulsar_timing}

As outlined above, obtaining pulse TOAs with \fm{} is fundamentally
different from the process traditionally followed in radio astronomy.
In the latter case, TOAs can be obtained with observations generally
lasting less than an hour, a time-scale on which TN is negligible.
The resulting pulse profiles follow Gaussian statistics even in the
low S/N case.  Measuring a TOA is then simply a matter of
cross-correlation with a known template.  In contrast, the LAT data
may comprise only a few source photons atop a strong background.  A
proper description of the data requires Poisson statistics, and
measuring a TOA is achieved through analysis of the likelihood
profile as we describe below.

\begin{figure}[t] 
\centerline{
\includegraphics[width=1.0\linewidth,clip=,angle=0]{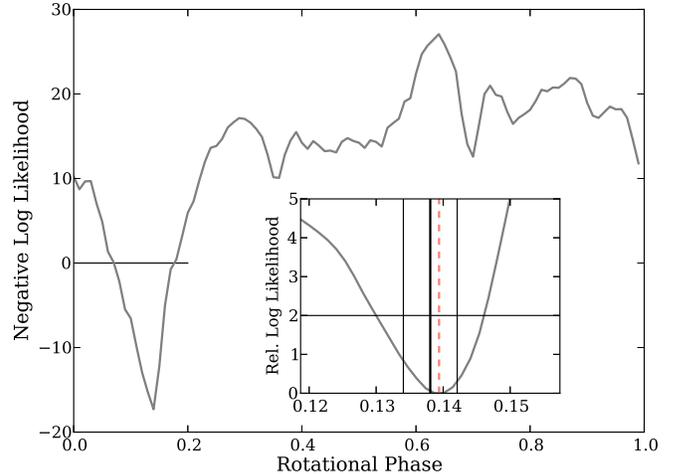}
}
\caption{\label{fig:logl}
The log likelihood as a function of relative phase between an analytic
template and a pulse profile obtained from two weeks of data for
PSR~J0030$+$0451.  There is a clear global extremum indicating a
well-measured TOA.  The log likelihood in the neighborhood of the
extremum (inset) is asymmetric.  As described in the text, instead of
the extremum (dashed salmon line), we adopt the center of the interval
defined by $\delta \log\mathcal{L}=2$ (solid black horizontal line).
This TOA, and the corresponding error bounds, are shown as black
vertical lines.
}
\end{figure}

In measuring TOAs, we follow the method of \citet{Ray11} with a few
refinements.  We use a weighted version of the likelihood, i.e.  the
probability density to observe a photon with phase $\phi$ given a
pulse template $f(\phi)$ is $p(\phi) = wf(\phi) + (1-w)$, where $w$ is
the photon weight computed above (\S\ref{sec:lat_reduction}).  For
TOAs with modest significance, the likelihood profile is often
asymmetric, indicating the statistical uncertainty on the resulting
TOA is also asymmetric.  However, standard pulsar timing software,
like the \textsc{Tempo2} package we use, is predicated on Gaussian
statistics and symmetric error bars.  Thus, instead of simply using
the likelihood maximum to measure a TOA, we determine a phase interval
bounded by $\delta\log \mathcal{L} = 2$ and we use the center of this
interval and one-half of its width as estimators for the TOA and its
uncertainty.  We show an example for PSR~J0030$+$0451 in Figure
\ref{fig:logl}.  Further, in order to avoid spurious TOA measurements,
we require that the absolute value of the log likelihood exceeds five,
a threshold we have determined empirically to work well.  If there are
multiple significant peaks in the likelihood, as is often the case for
profiles with peaks separated by about half a rotation, we choose the
peak closest to that predicted by the ephemeris.

We finally note that TOA extraction is an iterative process.
Determing the photon phase $\phi$ over a TOA interval requires a valid
timing solution, but this in turn requires a set of TOAs from which to
determine a timing solution.  We begin with an initial ephemeris that
accurately tracks the pulsar rotational phase for some span of time
(e.g. over three years from 2PC), and we extract TOAs using the method
of \citet{Ray11}.  The integration period for each TOA, which is
uniform for a particular pulsar, varies from one week for the
brightest pulsars in our sample to 56 days for the dimmest (see Tables
\ref{tab:overview_normal} and \ref{tab:overview_msp}).

Using the initial solution and TOAs, we jointly fit a new timing
solution and a model for timing noise using the Cholesky
factorization/generalized least squares method of C11 (see
\S\ref{sec:tn}).  Using the covariance matrix predicted by the
best-fit TN model, we evaluate the TN process in the time domain with
critical sampling (twice the TOA cadence).  See \citet{Deng12} for a
similar application.  With this new timing solution we extrapolate the
pulse phase to the next interval of data and extract a new TOA.  By
repeating this process for each time step, we bootstrap a self
consistent timing solution and set of TOAs.

We model glitches as sudden discontinuities in the spin-down model:
\begin{eqnarray}
\label{eq:glitch}
\phi(t) & = &\phi_0(t) + \theta(t-t_g)\times\nonumber\\
&\big\{&\phi_g + \delta\nu\,(t-t_g) +
\frac{1}{2}\delta\dot{\nu}\,(t-t_g)^2 + \nonumber\\
&&\delta\nu_d\tau_d\,[1-\exp\frac{-(t-t_g)}{\tau_d}]\big\},
\end{eqnarray}
where $t_g$ is the glitch epoch, $\theta$ is the Heaviside function,
$\delta\nu$ and $\delta\dot{\nu}$ are permanent changes in the
frequency and spin-down rate and $\delta\nu_d$ is a transient increase
in the frequency exponentially decaying with time-scale $\tau_d$.  New
glitches are relatively easy to identify, as the pre-glitch timing
solution will no longer produce a visible pulse profile when applied
to post-glitch data.  When we note such an occurrence, we estimate the
epoch and then use unbinned likelihood to estimate the glitch
parameters directly from the photon timestamps, fitting data collected
for 90\,d before and 180\,d after the glitch.  Empirically, we find
that a subset of these parameters suffice to model the set of
glitches.  However, we also note that (1) generally we cannot access
transient components with time-scales less than a few days (2)
parameter estimates are affected by TN.  Joint fitting of the glitch
parameters and the TN model is beyond the current capability of our
analysis and we leave this effort for future work, noting that the
impact of fitting the glitch parameters independently does not appear
to substantially alter the TN properties (see \S\ref{sec:spectrum}).

Once we have a timing solution spanning the full data set, we generate
a new pulse profile and fit an analytic pulse template as described in
2PC.  These templates are based on the 2PC models to which we have
added minor components to model new features present in our augmented
data set.  For consistency, we use this new pulse profile to
re-extract TOAs and fit a new timing solution.  Although this `self
standarding' can lead to underestimates of TOA uncertainty
\citep[e.g.][]{Hotan05}, we expect the effect to be minimal here
because (1) we use an analytic template (2) our fitting is in the time
domain.

\subsection{Data Products}
\label{sec:pipeline}

For each pulsar, the principal products of the pipeline are TOAs at an
optimal cadence, a timing solution including a model of TN, and a
pulse profile.  These products---along with diagnostic plots---are
public\footnote{http://slac.stanford.edu/\~{}kerrm/fermi\_pulsar\_timing}$^{,}$\footnote{http://fermi.gsfc.nasa.gov/ssc/data/access/lat/ephems}
and generally updated every few months.  The TN model, discussed
below, is characterized by critically sampling its time domain
realization (twice per TOA).  This sampling is encoded in the timing
solution via a set of \texttt{IFUNC} values, which specify the fixed
points of a time series between which \textsc{Tempo2} interpolates
linearly.  To compare these timing solutions with other published
ephemerides and light curves, note that $\phi=0$ in our timing
solutions corresponds to the leftmost bin of the corresponding
profile.

\section{Timing Noise}
\label{sec:tn}

\subsection{Maximum Likelihood Fitting}
\label{sec:ml}

\begin{figure*}
\centering
\plottwo{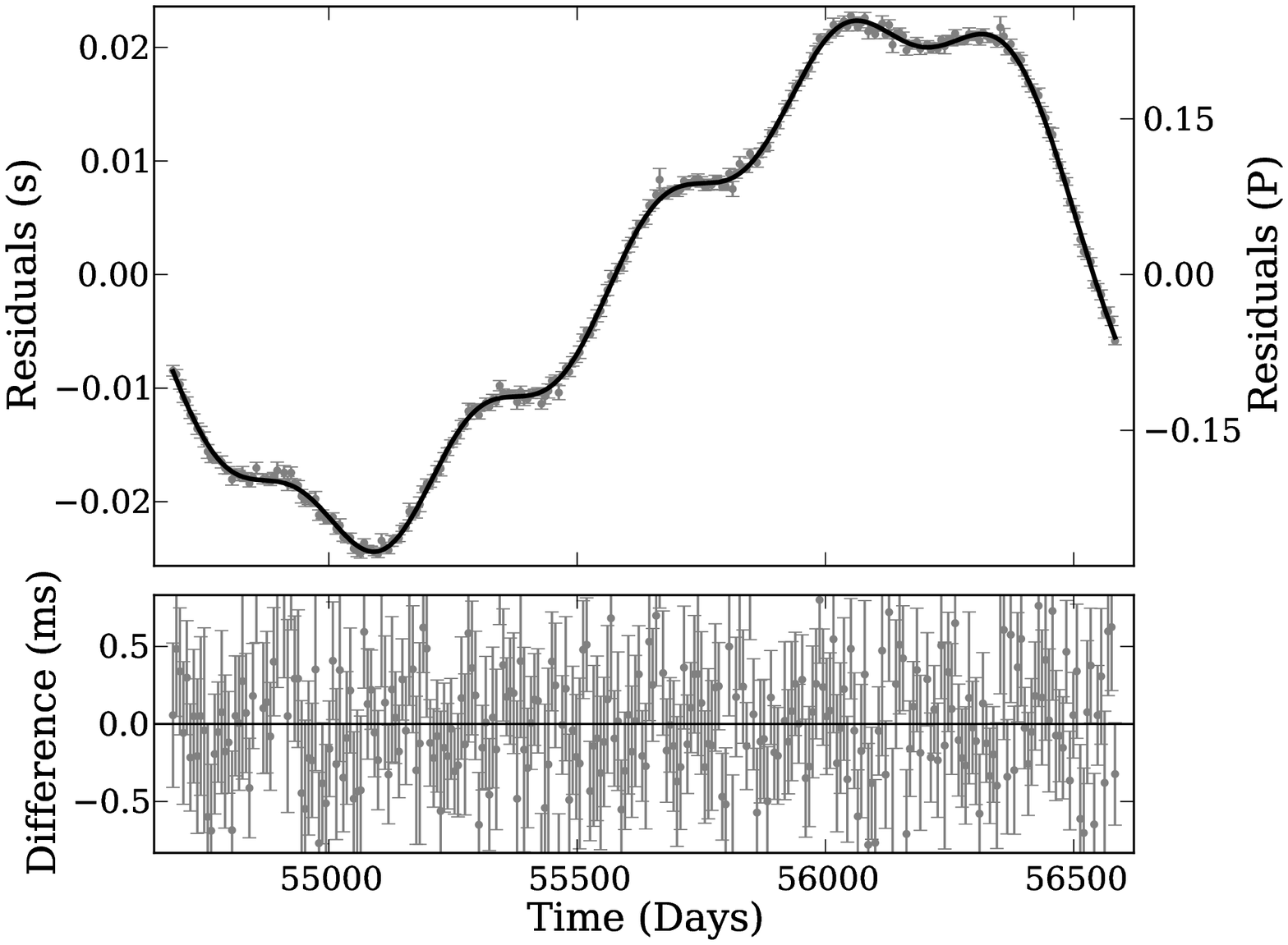}{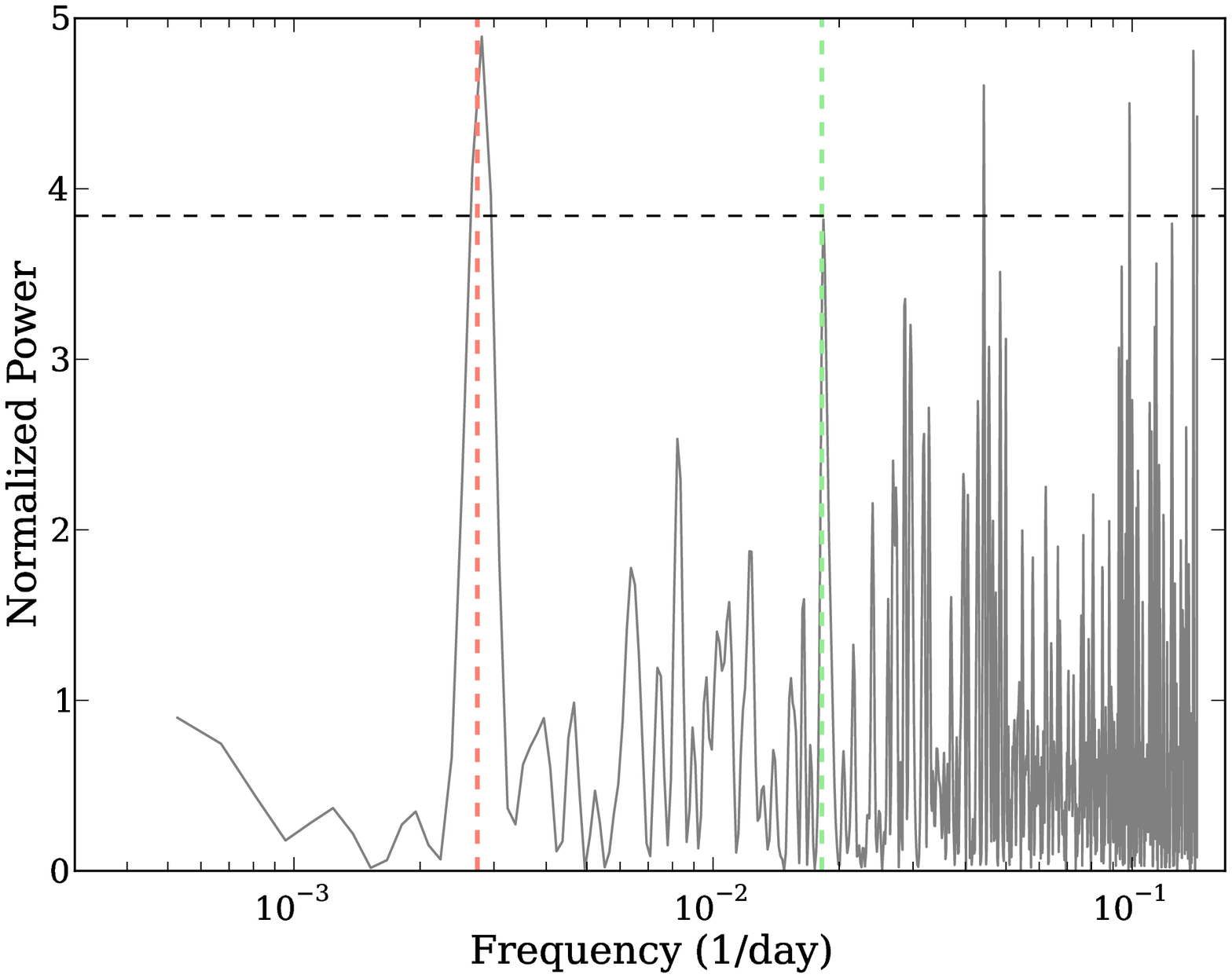}\\
\plottwo{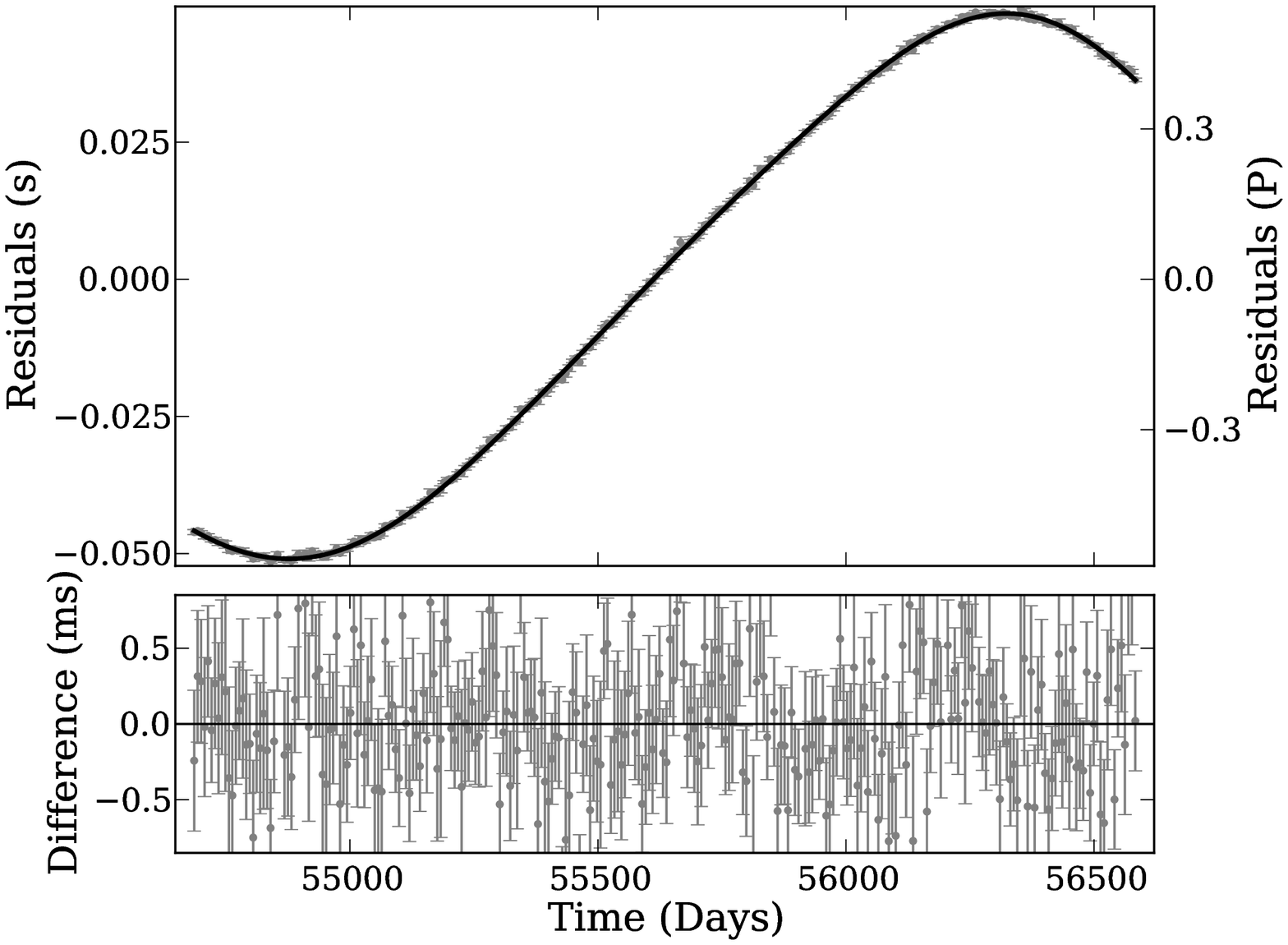}{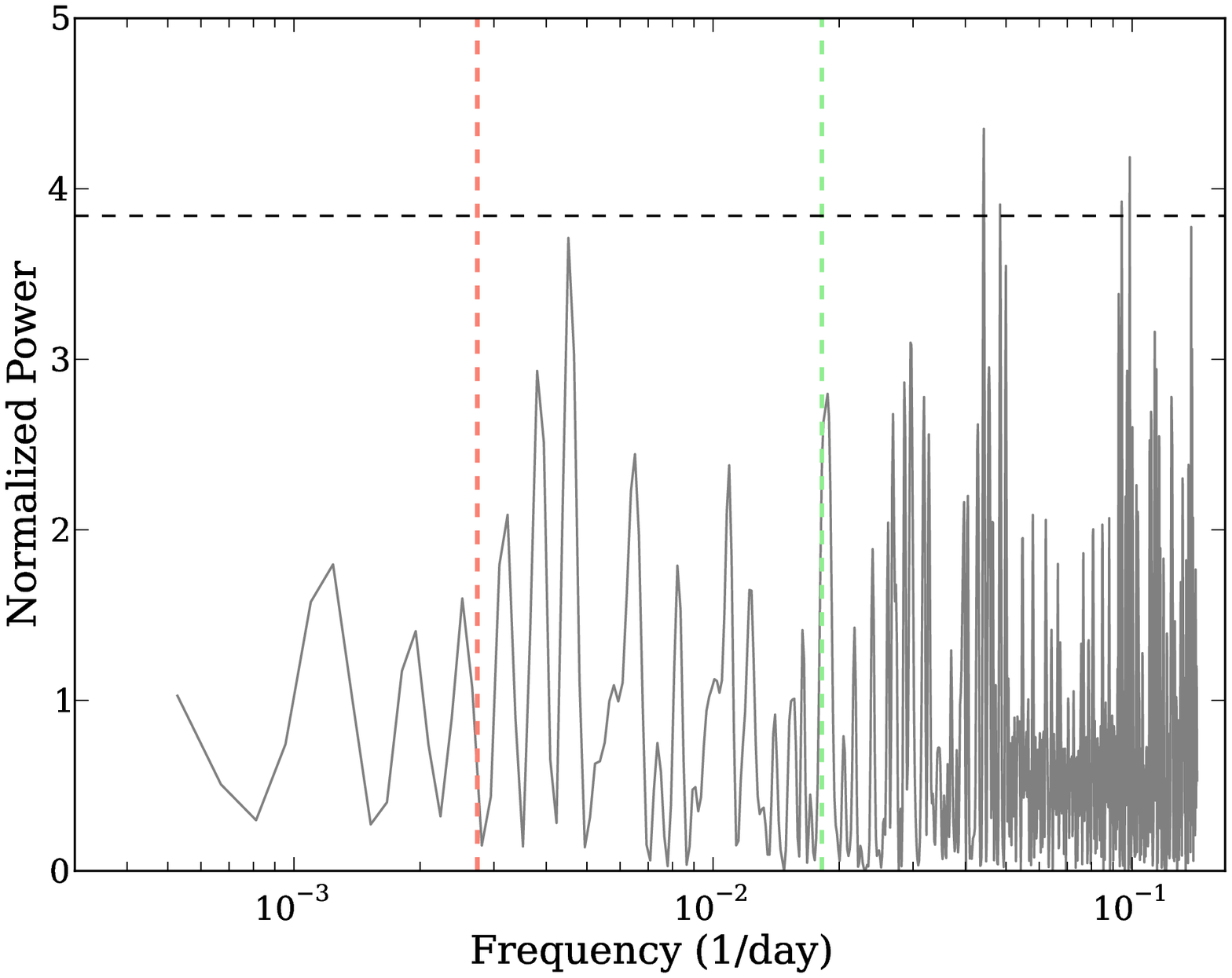}
\caption{\label{fig:fitting}
Example of the TN fitting process for PSR~J1028$-$5819.  The top
portions of the left-hand plots show the TOA residuals after a smooth
spin-down model has been subtracted, as well as the estimated
realization of the noise process (solid black line; see main text).
Note that the dominant noise term is quadratic (\nudot{}-like).  The
lower portions of the left-hand plots show the difference between the TOA
residuals and the noise estimator. The righthand plots show the power
spectrum (as realized by the Lomb-Scargle periodogram) of the
covariance-transformed residuals (see main text).  The dashed
horizontal line gives the amplitude that 5\% of independent samples
should surpass by chance, while the vertical salmon (green) line
indicates periods of 365 and 55 days.  In the top row, the position
has been fixed at the independent ATCA position \citet{Keith08}.  The
offset from the timing position is statistically significant and the
annual sinusoid appears superimposed on the realization of the red
noise and excess power at the annual scale is clearly present in the
periodogram.  In the second row, the same plots appear after a joint
fit of TN and position.  The TN model has steepened to account for the
reduced high frequency power, and accordingly the linear component
($\nu$-like) is strenghtened in the best-fit noise process.  The
annual power has vanished from the periodogram, which is largely
consistent with white noise.  Note the presence of aliasing in the
higher frequencies of the periodograms.}
\end{figure*}

Characterization of TN requires some model for the noise
process.  Historically, these models have included a Taylor expansion
of the red noise realization through \nudot{} and higher derivatives,
and/or through harmonically related sinusoids
\citep[\texttt{FITWAVES},][]{Hobbs04}.  With this approach, which
models the realization of the noise process in the time domain, the
measured TOAs remain statistically independent and maximum likelihood
analysis is equivalent to weighted least squares (i.e., the covariance
matrix is diagonal).

Unfortunately, C11 established that such ad hoc models of TN heavily
bias other timing parameters, and instead those authors propose that
the red noise process be modelled as widesense stationary, that is,
that the correlations between TOAs are independent of time.
Then, fitting can
occur via generalized least squares by minimizing $\chi^2\equiv
\mathbf{dx}^{\mathrm{T}}\,\mathbf{C}^{-1}\,\mathbf{dx}$, with
juxtaposition indicating matrix multiplication, $\mathbf{C\equiv
C_{meas} + C_{sto}}$, the combined covariance matrix of the white
measurement (diagonal) errors and stochastic process (nondiagonal,
symmetric), and $\mathbf{dx}$ being the residuals of the data with
respect to the spin-down model only.

If the covariance matrix $\mathbf{C_{sto}}$ is chosen accurately---by
which we mean the model reproduces the correlations observed in the
TOA time series---then
the above method provides optimal estimators for the timing
parameters.  C11 adopted a model in which the power spectrum of the TN
is described in the frequency domain as a power law
\begin{equation}
\label{eq:power_law}
S(f)=A^2 [1+(f/f_c)^2]^{-\alpha/2},
\end{equation}
with the ``corner frequency'' $f_c$ providing a low-frequency
saturation, $\alpha$ the spectral index, and $A$ the amplitude.   They
employed a spectral analysis of the residuals to the spin-down model
to estimate these three parameters.  The corner frequency was to some
extent empirical because their fit was to the residuals, from which a
quadratic term (viz. low frequency power) had already been removed by
fitting $\nu$ and \nudot{} \citep[see e.g.][]{VanHaasteren13}.  We
also adopt this model for TN, but as we are interested both in the
properties of the TN as well as its impact on other
parameters, we choose to fit the TN parameters jointly with the
spin-down model.  This requires retaining the normalization term in
the Gaussian likelihood; that is, denoting the stochastic TN
parameters $\lambda_{s}$ and the pulsar spin-down parameters
$\lambda_{p}$, we obtain parameter estimates by minimizing the
quantity
\begin{equation}
-2\log\mathcal{L}\equiv \log\det\mathbf{C}(\lambda_{s}) +
\mathbf{dx}^{\mathrm{T}}(\lambda_{p})\,\mathbf{C}^{-1}(\lambda_{s})\,\mathbf{dx}(\lambda_{p}).
\end{equation}
Moreover, we require the covariance matrix for the red noise process
to be positive definite, in which case the total covariance matrix is
necessarily positive definite, and we apply the Cholesky factorization
$\mathbf{C}=\mathbf{L}^{\mathrm{T}}\,\mathbf{L}$, with $\mathbf{L}$ a
lower triangular matrix.  Thus, $\log\det\mathbf{C}$ reduces to
$\sum_i \log L_{ii}$, and the ``$\chi^2$'' term can be written as
$(\mathbf{L}^{-1}\,\mathbf{dx})^{\mathrm{T}}\,(\mathbf{L}^{-1}\,\mathbf{dx})$.
For many pulsars with strong timing noise, the condition number of
$\mathbf{C_{sto}}$ is untenably large and the Cholesky factorization
greatly aids numerical stability.

To maximize $\log \mathcal{L}$ and characterize parameter
uncertainties, we use Monte Carlo Markov chain sampling via
\texttt{emcee} \citep{Foreman-Mackey13}.  For each proposed
realization of power law parameters $\lambda_s$, we compute the
covariance matrix for the $N\,(N-1)/2$ unique lags by computing the
Fourier transform of $S(f)$; see \citet{Perrodin13} for an analytic
expression of the transform and note that this is essentially the
well-known Mat\'{e}rn covariance function.  If the matrix is positive
definite, we factor it as described above.  Otherwise we reject the
sample.  Next, we use the proposed set of spin-down parameters
$\lambda_p$ together with \textsc{Tempo2} to compute
the residuals $\mathbf{dx}$.  Finally, we compute $\log \mathcal{L}$
and return it to \texttt{emcee} to let it determine whether or not to
accept the sample.

Because we have very little knowledge of the TN model \textit{a
priori}, we adopt broad priors on $\lambda_s$ by sowing the
\texttt{emcee} ``walkers'' over a parameter space encompassing a large
range of TN strengths and spectral shapes.  Specifically, we use a
prior uniform in the logarithm of $A^2$ from $-22$ to $12$, a uniform
prior on $\alpha$ from 2 to 10, and a uniform prior on the logarithm
of $f_c$ from $-5$ to $0$, with units on $A^2$ of s$^2$\,d and units
on $f_c$ of d$^{-1}$.  Conversely, the timing model parameters are
generally highly constrained, and we seed the walkers over a narrow
parameter space centered on values from a \textsc{Tempo2} fit with
width of order the formal statistical uncertainties from the ordinary
least squares fit.  We then allow a lengthy burn-in for the walkers to
``forget'' their initial conditions and explore the covariance between
$\lambda_s$ and $\lambda_p$.  As long as the posterior distribution
does not pile up on a boundary, we find our results are insensitive to
the volume of the parameter space.  We employ a basic spin-down model,
fitting only $\nu$ and \nudot{}, save for the Crab pulsar, where
\nudotdot{} is dynamically important and we also allow it to vary.
For other pulsars, we fix $\ddot{\nu}=3\,\dot{\nu}^2/\nu$, i.e.
assign a braking index of 3.  For Crab, we include $\dddot{\nu}$
likewise fixed by the braking index prescription.  If an
arcsecond-precision position is not available from {\it Chandra},
optical telescopes, or VLBI measurements (see \S\ref{sec:astrometry}),
or if the implied timing precision is much better than that of the
multi-wavelength position, we also allow the position to vary.  In
Figure \ref{fig:fitting}, we show one example where the systematic
errors associated with an ATCA position for PSR~J1028$-$5819 are
clearly revealed in the timing analysis.  Finally, the distribution of
samples from \texttt{emcee} provide estimates of uncertainty on
$\lambda$.  If we have included radio TOAs, we fit an additional
arbitrary phase offset between the \fm{} and radio TOAs. 

After this process, we have likelihood-maximizing parameters for the
spin-down and TN model, but we still lack a time-domain realization of
the noise process with which to generate a set of whitened residuals
and thus an optimal timing solution.  \citet{Deng12} derive a maximum
likelihood estimator for the noise process, but we find the required
matrix inversion to be numerically unstable for pulsars with strong
TN.  Instead, we once again adopt an iterative process.  According to
the Karhunen-Lo\`{e}ve theorem, the eigenvectors of $\mathbf{C_{sto}}$
are an optimal basis for an expansion of the noise process, and the
coefficients of the expansion can be estimated by ordinary least
squares to the timing residuals.  (See, e.g., \citet{Tegmark97} for an
overview of Karhunen-Lo\`{e}ve decomposition in an astronomical
context.)  We then approximate the noise process as a truncated
Karhunen-Lo\`{e}ve expansion, adding terms to the expansion until the
fit becomes acceptable, i.e. the $\chi^2$ per degree of freedom is
close to 1.  This procedure is similar to that of FITWAVES, but we use
a basis optimized to the noise process rather than a general harmonic
basis.  This realization of the noise process is included via linear
interpolation in the publicly-available ephemerides.

\subsection{Model Selection}
\label{sec:model_selection}

\begin{figure}[t] 
\centerline{
\includegraphics[width=1.0\linewidth,clip=,angle=0]{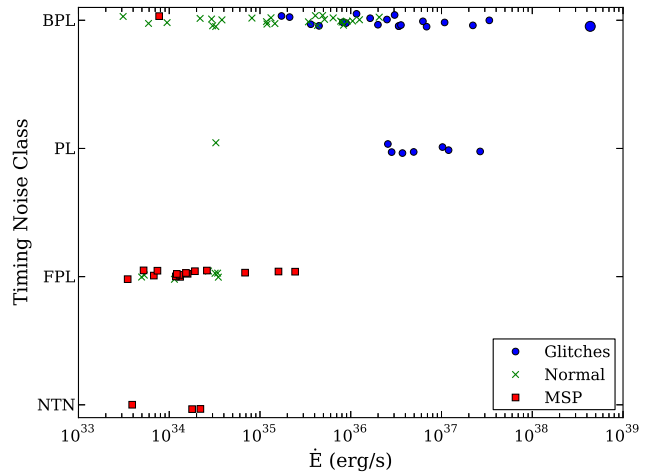}
}
\caption{\label{fig:comp_class_edot}
The timing noise model classification (see main text and Tables
\ref{tab:overview_normal} and \ref{tab:overview_msp}) of each pulsar:
Broken Power Law, (pure) Power Law, Fixed Power Law, and No Timing
Noise.  The points are staggered for visibility.  The legend
and symbols indicate three classes of pulsars: young pulsars having
experienced a glitch (``Glitches'', blue circles); young pulsars
without evidence glitches (``Normal'', green crosses); and millisecond
pulsars (``MSP'', red squares).  The symbol corresponding to the Crab
pulsar, J0534$+$2200, is drawn slightly enlarged here and in
subsequent figures.
}
\end{figure}

Although fitting $\nu$ and \nudot{} suppresses TN at time-scales
comparable to the observation length \citep[see][]{VanHaasteren13}, it
is interesting to see if the joint fitting process preserves any
evidence in support of noise processes with spectra well-described by
power laws.  We explore this by repeating the fitting process
described above for a series of nested models in which
\begin{enumerate}
\item $A=0$ in Eq. \ref{eq:power_law};  (no timing noise, `NTN'
in Tables \ref{tab:overview_normal} and \ref{tab:overview_msp});
\item $A>0$, $\alpha=3$, $f_c\ll1/T$ (pure power law with fixed spectral
index; `FPL' in Tables \ref{tab:overview_normal} and
\ref{tab:overview_msp});
\item $A>0$, $\alpha>0$, $f_c\ll1/T$ (pure power law with free
spectral index; `PL' in Tables \ref{tab:overview_normal} and
\ref{tab:overview_msp});
\item $A>0$, $\alpha>0$, $f_c>0$ (broken power law with low frequency
cutoff, `BPL' in Tables \ref{tab:overview_normal} and
\ref{tab:overview_msp}).
\end{enumerate}
$T$ is the
length of the data set, and the final case is precisely the fits
described in the previous sections.  Our choice of $\alpha=3$ represents red noise
of modest steepness as an alternative to the steeper red noise we
observe in the young pulsar sample. To select the optimal model, we
apply the Akaike information criterion \citep{Akaike73}, which requires
$\delta\log\mathcal{L}>2$ to accept a model with an additional degree
of freedom.  The resulting classification for each pulsar appears in
Tables \ref{tab:overview_normal} and \ref{tab:overview_msp}, and in
Figure \ref{fig:comp_class_edot}.

Every young pulsar provides evidence for at least some level of TN,
and for the vast majority, the preferred model is a broken power law.
Curiously, the pulsars for which a pure power law (case 3 above) is
optimal are mostly energetic, glitching pulsars.  Because the glitches
are not fit jointly with the TN or other spin parameters, they have a
complicated effect on the spectrum.  A data span $S$ (270\,d) is used
to determine the glitch parameters, and TN power with frequencies
$>$$1/S$ will be reduced by the glitch fitting.  On the other hand,
inaccuracies in the permanent glitch parameters ($\delta\nu$ and
$\delta\dot{\nu}$ in Eq. \ref{eq:glitch}) appear as a step with
asymptoptic spectrum $1/f^2$.  Since this is much less steep than the
intrinsic spectrum, however, the effect on the measured spectrum
should be minor.  Since pure power laws are preferred for glitching
pulsars, this seems to be the case.

With one exception, at most a pure power law with fixed spectral index
(case 2) is required to model the TN in MSPs.  In fact, since the
predicted red noise level (see below) is largely below the statistical
precision of our TOAs, it is likely most of this ``timing noise'' is
simply absorbing deficiencies in our white noise models, e.g. the
likelihood asymmetry described in \S\ref{sec:ml}.  This interpretation
is reinforced in the case of PSR~J1231$-$1411, which is the brightest
known \fm{} MSP, and thus least affected by issues with low
statistical significance.  Its optimal model requires no TN at all
(case 0).

\subsection{Measurements and Discussion}
\label{sec:measurements}

Substantial progress has been made in understanding the properties and
underlying mechanisms of timing noise through the study of large
samples of pulsars timed over many years,
\citep[e.g.][]{Cordes80b,Cordes85,DAlessandro95,Hobbs10}.  The
collection of pulsars timed here is comparable in both size and
duration to many of these studies.  Moreover, since the beams of
\g-ray pulsars illuminate more of the sky than those of radio pulsars
\citep[e.g.][]{Watters11},
the LAT sample may offer a less biased view of high-$\dot{E}$ pulsars.
Both circumstances motivate an analysis and discussion of two key TN
metrics, viz its strength and its spectral shape. 

\subsubsection{Amplitude}

\begin{figure}[t] 
\centerline{
\includegraphics[width=1.0\linewidth,clip=,angle=0]{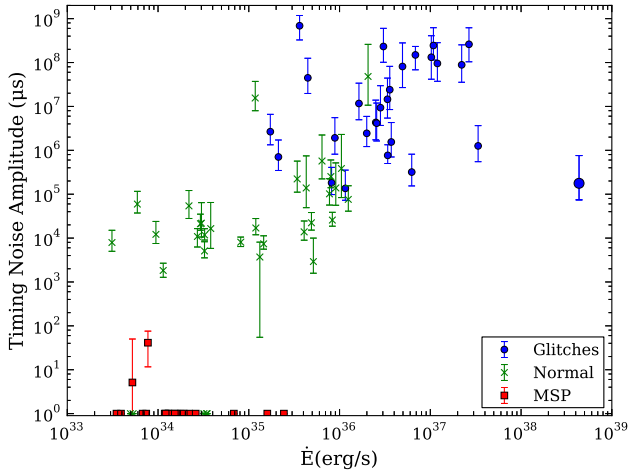}
}
\caption{\label{fig:tnedot}
The amplitude of the timing noise process as estimated from the
zero-lag autocorrelation of the best-fit TN model.  Normal
pulsars are classified according to those that do or do not suffer
glitches. The central values and error bars displayed here and in
subsequent figures with TN metrics are obtained from the central 68\%
of the Monte Carlo realizations described in the main text.  Pulsars
with undetectable levels of timing noise are shown without error bars
at 1\,$\mu$s amplitude.
}
\end{figure}

The strength of TN is a key parameter, as it sets the physical scale
of the underlying noise process(es) and limits the precision with
which deterministic parameters can be measured.  Several metrics are
available, including the best-fit value of $\ddot{\nu}$, the rms (root
mean square) after
fitting a low-order polynomial, and the $A$ parameter of Eq.
\ref{eq:power_law}.  Because the latter is highly correlated with
$f_c$, we adopt as a proxy for TN strength the integral timing noise
$\int S(f)\,df$ or, equivalently, the zero-lag element of the
time-domain covariance.  To account for the statistical uncertainty in
this quantity, we compute it using each entry of our Markov chain and
report the mode and 68\% central interval as a function of $\dot{E}$
in Figure \ref{fig:tnedot}.  As extensively noted in the literature,
as well as in our sample, the TN strength is highly correlated with
\nudot{} and mildly correlated with $\nu$, e.g.
\citep{Cordes80b,Arzoumanian94,Hobbs10}, though see also
\citet{Shannon10}.  For uniformity of presentation, we use $\dot{E}$
as a proxy for \nudot{}  and other combinations of $\nu$ and \nudot{}.
Interestingly, the Crab pulsar (J0534$+$2200) shows an anomalously low
TN amplitude.  Among young and energetic pulsars, it is unique in its
proclivity for small glitches with complex recoveries
\citep{Espinoza11}, and it is no surprise its TN properties are out of
family.  Moreover, the Crab is the only pulsar for which we fit
$\ddot{\nu}$.

\begin{figure}
\centerline{
\includegraphics[width=1.0\linewidth,clip=,angle=0]{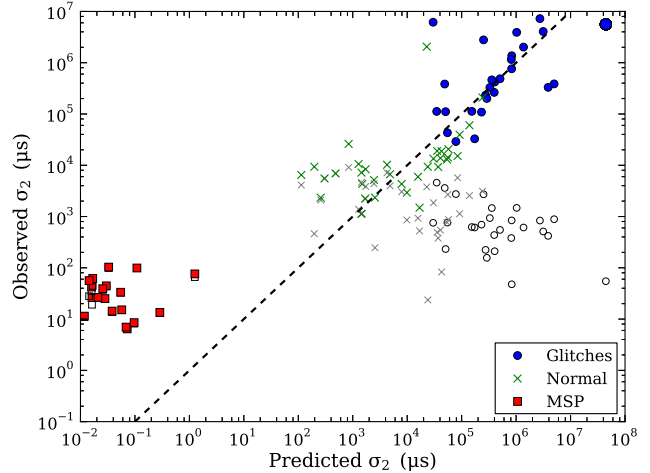}
}
\caption{\label{fig:rms_plot}
The residual scatter after modelling $\nu$ and \nudot{}, $\sigma_2$
(see main text).  The total scatter is depicted by larger colored
points, while below them, estimates for the sample rms with the
contribution of TN removed appear as smaller gray and white points.
The x-axis gives the prediction from the scaling relation of
\citet{Shannon10}, and the dashed line shows equality, i.e. agreement
with this model.  For pulsars with well-measured TN (colored points
well above white points), the observations are in reasonable
agreement.
}
\end{figure}

By examining a large sample of pulsars, \citet{Shannon10} identified a
single relation spanning MSPs, young pulsars, and magnetars:
$\sigma^2_2\propto\nu^{-1.8}\dot{\nu}^2$, where $\sigma_2$ is the rms
scatter of the TOA residuals after fitting for $\nu$ and \nudot{},
with the subscript reminding that the model includes two and only two
spin-down terms.  Measurements of $\sigma_2$ for the pulsar population
and the predictions of \citet{Shannon10} are shown in Figure
\ref{fig:rms_plot}; observations in agreement with the relation lie
along the diagonal dashed line.  These values are obtained by
measuring the rms scatter relative to a timing solution with $\nu$ and
\nudot{} and for which the best-fit noise process is either included
in the model (``white rms'') or excluded from the model (``total
rms'').  Generally, the total rms (colored points in Figure
\ref{fig:rms_plot}) dwarfs the white rms (white points) for young
pulsars, and the former can be taken as a proxy of the rms due to TN.
However, as measurement noise becomes important, the two quantities
approach equality, and we cannot reliably estimate the TN
contribution.  Although the values are in reasonable agreement with
the fit, the LAT sample shows a steeper dependence on \nudot, more in
keeping with the $\sigma^2_2\propto\nu^{-0.8}\dot{\nu}^{1.5}$ relation
found by \citet{Hobbs10b} in a fit to only typical unrecycled pulsars.
In both cases, the white noise level of the MSP sample is well above
the TN prediction.

\begin{figure}
\centerline{
\includegraphics[width=1.0\linewidth,clip=,angle=0]{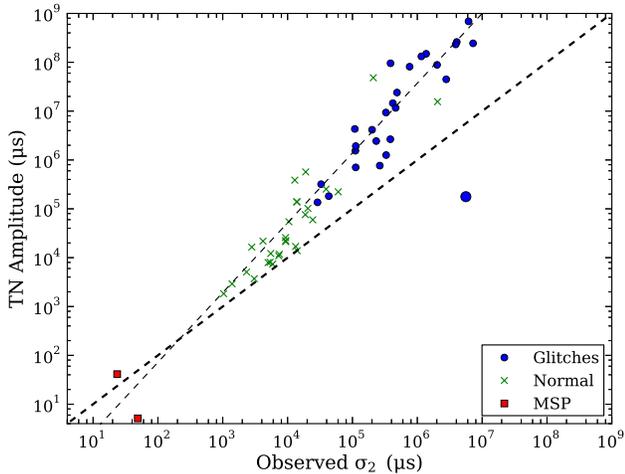}
}
\caption{\label{fig:comp_rms_ampl}
The timing noise strength, as in Figure \ref{fig:tnedot}, as a
function of the total rms, as per Figure \ref{fig:rms_plot}.  The
heavy dashed line indicates a linear relation, while the steeper,
lighter line has a (log log) slope of 7/5.  Pulsars with undetectable
timing noise are suppressed in this plot (see Figure
\ref{fig:tnedot}).}
\end{figure}

It is interesting to compare our model-derived metric (Figure
\ref{fig:tnedot}) with the sample rms in the case where we do not
model TN.  We expect that the primary difference will be the joint fit
of the TN model with $\nu$ and \nudot{}, potentially allowing more
noise to be associated with the TN model.  Indeed, in Figure
\ref{fig:comp_rms_ampl}, we observe a tight but nonlinear correlation,
with the TN predicted by our fit growing more quickly than the sample
rms.  The Crab pulsar lies more in trend with the $\sigma_2$ metric,
which likely reflects the inclusion of a dynamic \nudotdot{} in the fit
with which the TN amplitude is obtained.

\subsubsection{Spectrum}
\label{sec:spectrum}

\begin{figure}
\centerline{
\includegraphics[width=1.0\linewidth,clip=,angle=0]{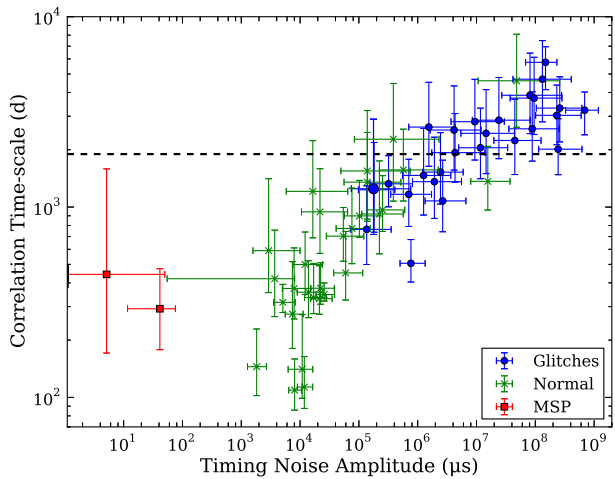}
}
\caption{\label{fig:comp_corr_tn}
The autocorrelation of the residuals (the Fourier transform of Eq.
\ref{eq:power_law}) monotonically decreases from zero lag.  Shown here
is the time required for the autocorrelation to drop to one half of
its peak value, plotted as a function of timing noise strength. The
black, dashed line indicates 5.2\,yr, the length of the data set, and
the clear trend appears to break modestly at this threshold.
}
\end{figure}

The second important property of TN---and the one that offers insight
into the physical mechanism(s)---is the degree of correlation over
long time-scales, or equivalently, the steepness of the power spectrum
of the TOA residuals.  Like the amplitude parameter, $\alpha$ is
highly correlated with $f_c$, and we prefer an empirical measure.  We
adopt the time required for the correlation to drop to one half of its
peak (zero lag) value, which we call the correlation time-scale, and
as before we make use of the Monte Carlo sampling to estimate values
and uncertainties.  These results appear in Figure
\ref{fig:comp_corr_tn} and show a very marked ``redder when stronger''
relation.  Discounting the MSPs, the relationship is nearly a power
law, though there is evidence for a break around 5\,yr, the length of
the data set, $T$.  Correlation time-scales $>T$ are of course
extrapolations and can be interpreted as statements that the
data exhibit strong correlations over the entire range of
observations.  Because we only have one realization of the data with
frequencies $\sim1/T$, and because this portion of the spectrum
dominates the total TN for steep spectra, precise characterization of
these quantities at these time-scales is not possible. 

\begin{figure}
\centerline{
\includegraphics[width=1.0\linewidth,clip=,angle=0]{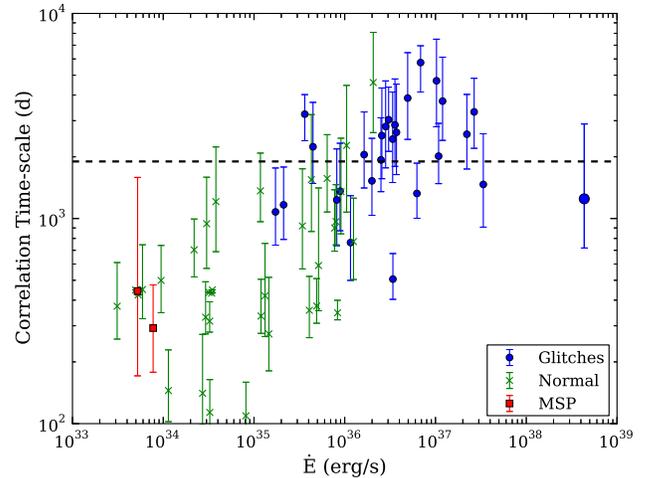}
}
\caption{\label{fig:corr_timescale_edot}As Figure
\ref{fig:comp_corr_tn} but with $\dot{E}$ for abscissa.  The
correlation time-scale increases monotonically with spin-down
luminosity, save for a possible saturation/break at the highest values
of $\dot{E}$.}
\end{figure}

Interestingly, though, this ``saturation'' time-scale is also the
time-scale at which most pulsars experience
strong glitches, so it is unclear if the saturation is due to glitches
or absorption of red noise by $\nu$ and \nudot{}.  Because the
correlation time-scale is so strongly correlated with the TN
amplitude, it is likewise strongly correlated with $\dot{E}$, as seen
in Figure \ref{fig:corr_timescale_edot}.

\begin{figure}
\centerline{
\includegraphics[width=1.0\linewidth,clip=,angle=0]{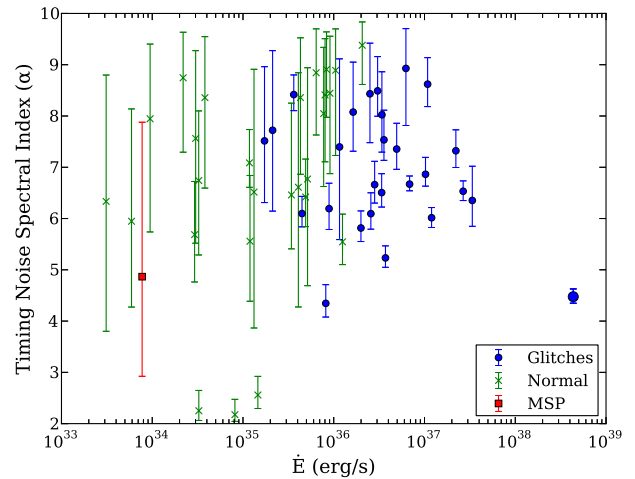}
}
\caption{\label{fig:alpha_edot}The TN spectral index, $\alpha$.  The
spectra are nearly universally steep, with values of 5--9, with little
evidence for dependence on $\dot{E}$.}
\end{figure}

As with TN strength, it is worth comparing our choice of the correlation
time-scale with other metrics of the TN spectrum.  The obvious choice,
$\alpha$, is directly comparable to spectral analyses based on Fourier
transforms, though we note the method employed here is immune to
spectral leakage and we easily recover extremely steep spectra:  our
measurements of $\alpha$ appear in Figure \ref{fig:alpha_edot}.
However, because of the correlation with $f_c$, $\alpha$ does not tell
the full story  and the strong correlation with $\dot{E}$ has vanished.
Indeed, the most energetic pulsars may show a decrease in spectral index!  In the time domain, on the other hand, the correlation
increases monotonically with $\dot{E}$.  We thus caution that, in
studies of TN of young pulsars, careful attention should be paid to
the metric employed.

On the other hand, the use of $\alpha$ allows us to make better
contact with the literature.  In the earliest studies of timing noise,
it was realized that (under the assumption that the `step
rate' is fast enough), simple random walk processes would manifest as
power law spectra.  In particular, for a random walk in phase,
$\alpha=2$; for $\nu$, $\alpha=4$; and for \nudot{}, $\alpha=6$.
In a famous example, \citet{Boynton72} determined the spectral index of the Crab pulsar,
based on two years of timing data, to be $\approx 4$, consistent with
our results here, and proposed its noise was largely `frequency
jitter'.  However, the majority of our pulsars show much steeper
spectra, with typical values of 6--7 for the most energetic,
glitching pulsars, and a broad range 5--9 for less energetic glitching
and non-glitching young pulsars.  As noted before, the TN properties
of the MSPs in our sample are poorly constrained.

We conclude this section with a few remarks on the interplay between
glitches and the (possibly) smooth noise process(es).  In the case of
large, resolved glitches, we expect we may be able to cleanly separate
the effects \citep[e.g.][]{Alpar86}, and indeed we observe no
systematic differences between the glitching and non-glitching
populations save the `normal' evolution with $\dot{E}$ and a tendency
to favor pure power laws.  On the other
hand, while \citet{Espinoza14} clearly show a minimum threshold for
glitch amplitudes of the Crab pulsar, there is no corresponding result
for the general population, and glitch amplitudes may stretch well
into the `measurement noise' resulting from both TOA precision and
cadence.  Indeed, \citet{Cordes85} found `microglitches' to be an
important factor in TN.  These episodic jumps in $\nu$ and \nudot{}
were too large to be consistent with high-rate stochastic processes
but were too small and with the incorrect sign for $\delta\nu$ to be
true glitches.  Such events are of particular concern for the LAT:
they may be the hallmark of mode changes \citep{Allafort13}, but the
typical TOA precision and cadence are poor for many pulsars.  A
thorough analysis of this phenomenon requires new capabilities to
jointly model glitches and TN, as well as to determine the sensitvity
of \fm{} to the former.  We leave this challenging task to future
work.

\section{Astrometry}
\label{sec:astrometry}

Pulsar timing can furnish extremely precise positions relative to the
dynamical frame of the Solar System ephemeris, in this work typically
realized by JPL DE405.  For PSRs~J0437$-$4715 and J2241$-$5236, we use
the DE414 ephemeris, while for PSRs~J1057$-$5226 and J2043$+$1711 we
make use of the DE421 ephemeris.  While small errors from frame
transfer or from clock drifts are important for MSP positions, the
dominant source of error for young pulsars is bias from TN
\citep[see e.g.][]{Weisskopf11,Ray11}.  For multi-wavelength
followup---e.g. deep searches for radio pulsations or X-ray searches
for point sources---accurate positions with robust uncertainties are
required.

Here, we show that our models of TN, together with Monte Carlo
sampling (described above), provide the needed level of accuracy.  To
estimate the timing position uncertainties, we compute the Monte Carlo
sample covariance in R.A. and Decl., obtaining the position error
ellipse marginalized over all other parameters.  Typically,
correlation in R.A. and Decl. is small, and we simply report the 68\%
(``1\,$\sigma$'') containment intervals for each coordinate.

To check these uncertainty estimates in the case of normal pulsars
(affected by TN), we have collected from the literature all available
precisely-measured pulsar positions, primarily from \textit{Chandra}
observations, but including some VLBI measurements and optical
observations.  We have intentionally excluded positions available from
pulsar timing as they are subject to the same potential biases as our
analysis.  The positions, along with their provenance and uncertainty,
are given in Tables \ref{tab:overview_normal} and
\ref{tab:overview_msp}.  The \textit{Chandra} position uncertainties,
in particular, are heterogeneous, as the level of analysis ranges from
none to a careful accounting of uncertainties due to photon counting,
pointing aspect solution, and frame tie.  For MSPs, the only positions
with sufficient precision to be useful for comparison are obtained
with VLBI, and we do not report positions with lesser precision.

\begin{figure}
\centerline{
\includegraphics[width=1.0\linewidth,clip=,angle=0]{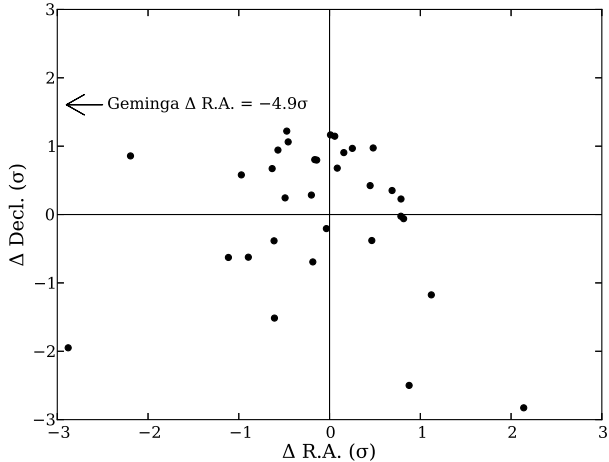}
}
\caption{\label{fig:pulls}
The difference between the timing and multi-wavelength positions of
Tables \ref{tab:overview_normal} and \ref{tab:overview_msp} in
``$\sigma$'' units, i.e. the timing and multi-wavelength errors summed
in quadrature (see main text).
}
\end{figure}

To compare the timing positions with the independent literature
positions, we referenced all timing positions to MJD 55555, while, for
sources with reliable proper motions, we advanced the independent
positions to the same epoch.  The resulting coordinate differences,
scaled to $\sigma^2=\sigma^2_{\mathrm{LAT}} +
\sigma^2_{\mathrm{MWL}}$, i.e. the errors added in quadrature, are
shown in Figure \ref{fig:pulls}.  The errors cluster within one unit
without piling up at the origin, indicating that the timing position
ellipse overlaps the multi-wavelength one without being overly
conservative.  Likewise, the \fm{} data indicate no additional
component of proper motion to within our measurement precision.

The one significant outlier is Geminga (PSR~J0633$+$1746), whose Decl.
is in agreement but whose R.A. differs by 5$\sigma$.  Here, its
absolute R.A. is taken from the \textit{Hubble} and \textit{Hipparcos}
measurements of \citet{Caraveo98} as
06$^{\mathrm{h}}$33$^{\mathrm{m}}$54$\fs$1530$\pm$0$\fs$0028 with an
epoch of MJD 49793.5, while the angular proper motion
$\mu_{\alpha}\cos\delta$ is measured with \textit{Hubble} by
\citet{Faherty07} as $142.2\pm1.1$\,mas/yr.  At the \fm{} epoch MJD
55555, the total change in angle is $2\farcs245\pm0\farcs017$, or a
coordinate change of $0\fs157\pm0\fs001$, giving a final position
06$^{\mathrm{h}}$33$^{\mathrm{m}}$54$\fs$310$\pm$0$\fs$003.  This
differs from the \fm{} position by $0\fs022$, or $0\farcs31$.
Accounting for the discrepancy requires either a 15\% systematic error
on the measured proper motion, or a systematic error of 300\,mas
in the optical frame tie; the latter seems more likely.

\begin{figure}
\centerline{
\includegraphics[width=1.0\linewidth,clip=,angle=0]{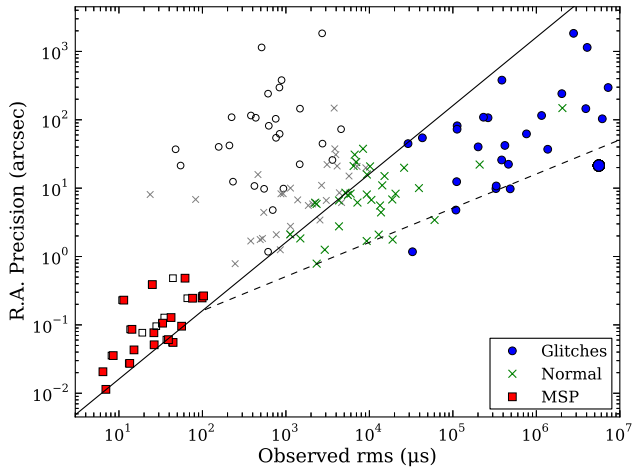}
}
\caption{\label{fig:comp_rms_raj}
The estimated precision on R.A. as a function of the observed scatter,
as per Figure \ref{fig:rms_plot}, save that here we indicate the
``white noise'' rms with gray, unfilled smaller markers, while the
colored larger ones give the sample rms.  The solid line indicates a
linear relationship and indicates the best R.A. precision possible for
the given white noise level.  The dashed line follows
$\sqrt{\mathrm{rms}}$.}
\end{figure}

The statistical uncertainty estimates for our positions are clearly
sufficiently accurate.  Moreover, we can limit the absolute level of
any systematic error within our timing chain itself by comparing the
\fm{}-only timing position of MSPs with precise VLBI positions.  In
the case of PSR~J0437$-$4715, the timing position is
in perfect agreeement with the 1-mas precision VLBI position of
\citet{Deller08}, limiting any systematic errors to $<$10\,mas.
Several \fm{}-LAT MSP positions have $\sim$mas precision, e.g.
J0614$-$3329 and J1231$-$1411 (and see Figure \ref{fig:comp_rms_raj}),
but those sources currently lack VLBI positions for comparison.

Although the positions above are unbiased, it is interesting to
explore the degree to which TN degrades the possible precision.  In
Figure \ref{fig:comp_rms_raj}, we show the estimated precision in the
R.A. coordinate as a function of both the total and white-only rms.
In general, the MSPs and a few of the less energetic young pulsars
track the ideal relation of
$\sigma_{\mathrm{R.A.}}\propto\sigma_{\mathrm{TOA}}$.  On the other
hand, when TN becomes important, the position precision worsens by
orders of magnitude.  Interestingly, if we instead examine the
precision as a function of the sample rms, which includes the effects
of TN, we see that the position uncertainty is still highly correlated
with the rms, but with an approximate
$\sigma_{\mathrm{R.A.}}\propto\sqrt{\mathrm{rms}}$ relation.
Essentially, the TN has destroyed the coherence of the position
measurement.  Thus, in the TN-dominated r\'{e}gime, we expect the
position uncertainty to only improve as $1/\sqrt{t}$, while those
pulsars free of TN (relative to the TOA precision) will
enjoy a $1/t$ improvement in localization.  We note that the
best-measured positions of young pulsars have precisions of
$<$100\,mas, e.g. J1028$-$5819, suggesting that proper motion
measurements may be feasible with longer data sets.

\begin{figure} 
\centerline{
\includegraphics[width=0.9\linewidth,bb=30 100 548 642,clip=,angle=0]{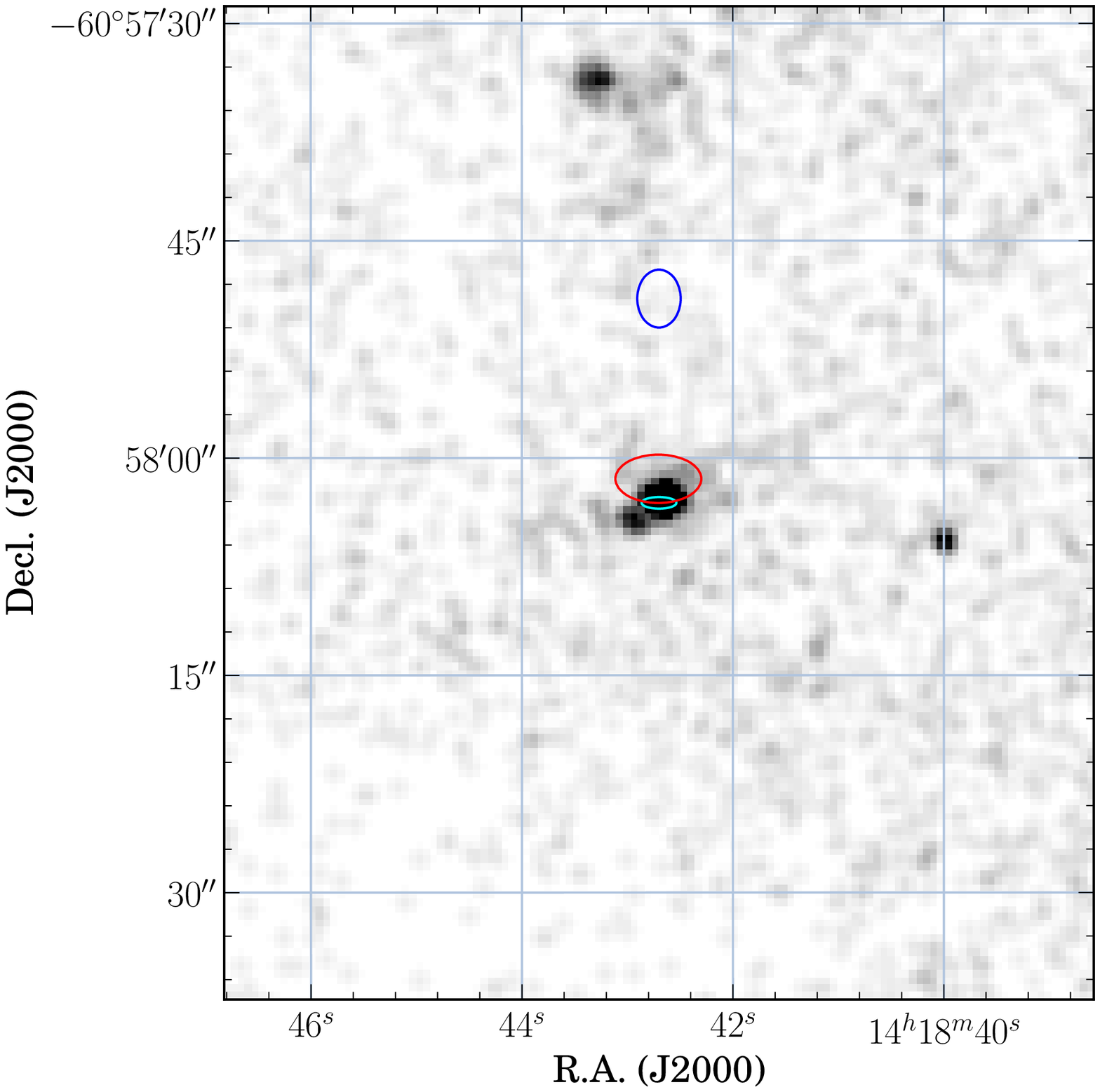}
}
\caption{\label{fig:rabbit}
The timing position of the radio-quiet PSR~J1418$-$6058---given by the
larger, red ellipse---agrees well with the \textit{Chandra} X-ray
position obtained by \citet{Ng05}, shown as the smaller, cyan ellipse.
The timing position of \citet{Ray11} is shown with the blue ellipse
offset to the north of the \textit{Chandra} source.  The archival
\textit{Chandra} image (obs. 7640) has been smoothed with a Gaussian
kernel of two pixels and the pointing aspect has not been
corrected.
}
\end{figure}

We conclude with a case study of the power of robust position
measurements.  A prior attempt by \citet{Ray11} to identify the X-ray
counterpart of PSR~J1418$-$6058, a radio-quiet pulsar embedded in the
``Rabbit'' pulsar wind nebula, could not discriminate between two
\textit{Chandra} point sources dubbed ``R1'' and ``R2'' \citep{Ng05}.
Although their timing position had sufficient formal statistical
precision, its centroid was substantially biased by TN and consistent
with neither X-ray source.  With appreciably more data and a reliable
TN model, we achieve a position uncertainty of $<$1\arcsec\ and can
firmly identify ``R1'' as the X-ray counterpart (Figure
\ref{fig:rabbit}).

\section{Summary and Conclusions}

We have presented results of timing a large sample of both radio-quiet
and radio-loud $\gamma$-ray pulsars with a long span of \fm{} data.
The strong TN in young pulsars required development of a new method in
which a model for the stochastic TN process is fit jointly with the
pulsar parameters, allowing characterization of the former and
unbiased estimation of the latter.  By comparing our best-fit timing
positions with precise results from the literature, we demonstrated
the efficacy of the method and were able to resolve a longstanding
ambiguity in the X-ray counterpart of PSR~J1418$-$6058.  Our study of
the TN properties indicates the spectra are generally quite steep and
may reflect an unmodelled glitch contributions---e.g. recoveries from
glitches both preceding and during our acquisition, or glitches below
detection thresholds, i.e.  `microglitches'.

The methods presented here are effective for more than half of the
\fm{} pulsar sample.  Newly detected pulsars, however, will
predominantly be fainter than those analyzed here and may not be
amenable to TOA extraction.  We hope the interesting results from this
work spur the community to develop the techniques required for
``photon-based'' pulsar timing which would allow any $\gamma$-ray
pulsar to be so characterized.



\acknowledgments

The Parkes radio telescope is part of the Australia Telescope, which
is funded by the Commonwealth Government for operation as a National
Facility managed by CSIRO.

The Robert C. Byrd Green Bank Telescope (GBT) is operated by the
National Radio Astronomy Observatory, a facility of the National
Science Foundation operated under cooperative agreement by Associated
Universities, Inc.

The \textit{Fermi} LAT Collaboration acknowledges generous ongoing
support from a number of agencies and institutes that have supported
both the development and the operation of the LAT as well as
scientific data analysis.  These include the National Aeronautics and
Space Administration and the Department of Energy in the United
States, the Commissariat \`a l'Energie Atomique and the Centre
National de la Recherche Scientifique / Institut National de Physique
Nucl\'eaire et de Physique des Particules in France, the Agenzia
Spaziale Italiana and the Istituto Nazionale di Fisica Nucleare in
Italy, the Ministry of Education, Culture, Sports, Science and
Technology (MEXT), High Energy Accelerator Research Organization (KEK)
and Japan Aerospace Exploration Agency (JAXA) in Japan, and the
K.~A.~Wallenberg Foundation, the Swedish Research Council and the
Swedish National Space Board in Sweden.
 
Additional support for science analysis during the operations phase
from the following agencies is also gratefully acknowledged: the
Istituto Nazionale di Astrofisica in Italy and the Centre National
d'\'Etudes Spatiales in France.

\bibliographystyle{apj}

\begin{thebibliography}{}

\bibitem[{{Abdo} {et~al.}(2010){Abdo}, {Ackermann}, {Ajello}, {Baldini},
  {Ballet}, {Barbiellini}, {Bastieri}, {Baughman}, {Bechtol}, {Bellazzini},
  {Berenji}, {Blandford}, {Bloom}, {Bonamente}, {Borgland}, {Bregeon}, {Brez},
  {Brigida}, {Bruel}, {Burnett}, {Buson}, {Caliandro}, {Cameron}, {Camilo},
  {Caraveo}, {Casandjian}, {Cecchi}, {{\c C}elik}, {Chekhtman}, {Cheung},
  {Chiang}, {Ciprini}, {Claus}, {Cognard}, {Cohen-Tanugi}, {Cominsky},
  {Conrad}, {Cutini}, {de Angelis}, {de Palma}, {Digel}, {Dingus}, {Dormody},
  {Silva}, {Drell}, {Dubois}, {Dumora}, {Farnier}, {Favuzzi}, {Fegan}, {Focke},
  {Fortin}, {Frailis}, {Freire}, {Fukazawa}, {Funk}, {Fusco}, {Gargano},
  {Gasparrini}, {Gehrels}, {Germani}, {Giavitto}, {Giebels}, {Giglietto},
  {Giordano}, {Glanzman}, {Godfrey}, {Grenier}, {Grondin}, {Grove},
  {Guillemot}, {Guiriec}, {Hanabata}, {Harding}, {Hays}, {Hughes}, {Jackson},
  {J{\'o}hannesson}, {Johnson}, {Johnson}, {Johnson}, {Johnston}, {Kamae},
  {Katagiri}, {Kataoka}, {Kawai}, {Kerr}, {Kn{\"o}dlseder}, {Kocian}, {Kuss},
  {Lande}, {Latronico}, {Lemoine-Goumard}, {Longo}, {Loparco}, {Lott},
  {Lovellette}, {Lubrano}, {Makeev}, {Marelli}, {Mazziotta}, {McEnery},
  {Meurer}, {Michelson}, {Mitthumsiri}, {Mizuno}, {Moiseev}, {Monte},
  {Monzani}, {Morselli}, {Moskalenko}, {Murgia}, {Nolan}, {Norris}, {Nuss},
  {Ohsugi}, {Omodei}, {Orlando}, {Ormes}, {Paneque}, {Parent}, {Pelassa},
  {Pepe}, {Pesce-Rollins}, {Piron}, {Porter}, {Rain{\`o}}, {Rando}, {Ray},
  {Razzano}, {Reimer}, {Reimer}, {Reposeur}, {Ritz}, {Roberts}, {Rochester},
  {Rodriguez}, {Ro'mani}, {Roth}, {Ryde}, {Sadrozinski}, {Sanchez}, {Sander},
  {Saz Parkinson}, {Scargle}, {Sgr{\`o}}, {Siskind}, {Smith}, {Smith},
  {Spandre}, {Spinelli}, {Strickman}, {Suson}, {Tajima}, {Takahashi}, {Tanaka},
  {Thayer}, {Thayer}, {Theureau}, {Thompson}, {Tibaldo}, {Tibolla}, {Torres},
  {Tosti}, {Tramacere}, {Uchiyama}, {Usher}, {Van Etten}, {Vasileiou},
  {Venter}, {Vilchez}, {Vitale}, {Waite}, {Wang}, {Watters}, {Winer}, {Wolff},
  {Wood}, {Ylinen}, \& {Ziegler}}]{Abdo10_1907}
{Abdo}, A.~A., {Ackermann}, M., {Ajello}, M., {et~al.} 2010, \apj, 711, 64

\bibitem[{{Abdo} {et~al.}(2013){Abdo}, {Ajello}, {Allafort}, {Baldini},
  {Ballet}, {Barbiellini}, {Baring}, {Bastieri}, {Belfiore}, {Bellazzini}, \&
  et~al.}]{Abdo13_2pc}
{Abdo}, A.~A., {Ajello}, M., {Allafort}, A., {et~al.} 2013, \apjs, 208, 17

\bibitem[{Akaike(1998)}]{Akaike73}
Akaike, H. 1998, in Selected Papers of Hirotugu Akaike (Springer), 199--213

\bibitem[{{Allafort} {et~al.}(2013){Allafort}, {Baldini}, {Ballet},
  {Barbiellini}, {Baring}, {Bastieri}, {Bellazzini}, {Bonamente}, {Bottacini},
  {Brandt}, {Bregeon}, {Bruel}, \& et~al.}]{Allafort13}
{Allafort}, A., {Baldini}, L., {Ballet}, J., {et~al.} 2013, \apjl, 777, L2

\bibitem[{{Alpar} {et~al.}(1986){Alpar}, {Nandkumar}, \& {Pines}}]{Alpar86}
{Alpar}, M.~A., {Nandkumar}, R., \& {Pines}, D. 1986, \apj, 311, 197

\bibitem[{{Arzoumanian} {et~al.}(1994){Arzoumanian}, {Nice}, {Taylor}, \&
  {Thorsett}}]{Arzoumanian94}
{Arzoumanian}, Z., {Nice}, D.~J., {Taylor}, J.~H., \& {Thorsett}, S.~E. 1994,
  \apj, 422, 671

\bibitem[{{Atwood} {et~al.}(2009){Atwood}, {Abdo}, {Ackermann}, {Althouse},
  {Anderson}, {Axelsson}, {Baldini}, {Ballet}, {Band}, {Barbiellini},
  {Bartelt}, {Bastieri}, {Baughman}, {Bechtol}, {B{\'e}d{\'e}r{\`e}de},
  {Bellardi}, {Bellazzini}, {Berenji}, {Bignami}, \& et~al.}]{Atwood09}
{Atwood}, W.~B., {Abdo}, A.~A., {Ackermann}, M., {et~al.} 2009, \apj, 697, 1071

\bibitem[{{Boynton} {et~al.}(1972){Boynton}, {Groth}, {Hutchinson}, {Nanos},
  {Partridge}, \& {Wilkinson}}]{Boynton72}
{Boynton}, P.~E., {Groth}, E.~J., {Hutchinson}, D.~P., {et~al.} 1972, \apj,
  175, 217

\bibitem[{{Brisken} {et~al.}(2003){Brisken}, {Thorsett}, {Golden}, \&
  {Goss}}]{Brisken03}
{Brisken}, W.~F., {Thorsett}, S.~E., {Golden}, A., \& {Goss}, W.~M. 2003,
  \apjl, 593, L89

\bibitem[{{Camilo} {et~al.}(2012{\natexlab{a}}){Camilo}, {Ransom},
  {Chatterjee}, {Johnston}, \& {Demorest}}]{Camilo12}
{Camilo}, F., {Ransom}, S.~M., {Chatterjee}, S., {Johnston}, S., \& {Demorest},
  P. 2012{\natexlab{a}}, \apj, 746, 63

\bibitem[{{Camilo} {et~al.}(2006){Camilo}, {Ransom}, {Gaensler}, {Slane},
  {Lorimer}, {Reynolds}, {Manchester}, \& {Murray}}]{Camilo06}
{Camilo}, F., {Ransom}, S.~M., {Gaensler}, B.~M., {et~al.} 2006, \apj, 637, 456

\bibitem[{{Camilo} {et~al.}(2004){Camilo}, {Manchester}, {Lyne}, {Gaensler},
  {Possenti}, {D'Amico}, {Stairs}, {Faulkner}, {Kramer}, {Lorimer},
  {McLaughlin}, \& {Hobbs}}]{Camilo04}
{Camilo}, F., {Manchester}, R.~N., {Lyne}, A.~G., {et~al.} 2004, \apjl, 611,
  L25

\bibitem[{{Camilo} {et~al.}(2009){Camilo}, {Ray}, {Ransom}, {Burgay},
  {Johnson}, {Kerr}, {Gotthelf}, {Halpern}, {Reynolds}, {Romani}, {Demorest},
  {Johnston}, {van Straten}, {Saz Parkinson}, {Ziegler}, {Dormody}, {Thompson},
  {Smith}, {Harding}, {Abdo}, {Crawford}, {Freire}, {Keith}, {Kramer},
  {Roberts}, {Weltevrede}, \& {Wood}}]{Camilo09}
{Camilo}, F., {Ray}, P.~S., {Ransom}, S.~M., {et~al.} 2009, \apj, 705, 1

\bibitem[{{Camilo} {et~al.}(2012{\natexlab{b}}){Camilo}, {Kerr}, {Ray},
  {Ransom}, {Johnston}, {Romani}, {Parent}, {DeCesar}, {Harding}, {Donato},
  {Saz Parkinson}, {Ferrara}, {Freire}, {Guillemot}, {Keith}, {Kramer}, \&
  {Wood}}]{Camilo12b}
{Camilo}, F., {Kerr}, M., {Ray}, P.~S., {et~al.} 2012{\natexlab{b}}, \apj, 746,
  39

\bibitem[{{Caraveo} {et~al.}(1998){Caraveo}, {Lattanzi}, {Massone}, {Mignani},
  {Makarov}, {Perryman}, \& {Bignami}}]{Caraveo98}
{Caraveo}, P.~A., {Lattanzi}, M.~G., {Massone}, G., {et~al.} 1998, \aap, 329,
  L1

\bibitem[{{Coles} {et~al.}(2011){Coles}, {Hobbs}, {Champion}, {Manchester}, \&
  {Verbiest}}]{Coles11}
{Coles}, W., {Hobbs}, G., {Champion}, D.~J., {Manchester}, R.~N., \&
  {Verbiest}, J.~P.~W. 2011, \mnras, 418, 561

\bibitem[{{Cordes} \& {Downs}(1985)}]{Cordes85}
{Cordes}, J.~M., \& {Downs}, G.~S. 1985, \apjs, 59, 343

\bibitem[{{Cordes} \& {Greenstein}(1981)}]{Cordes81}
{Cordes}, J.~M., \& {Greenstein}, G. 1981, \apj, 245, 1060

\bibitem[{{Cordes} \& {Helfand}(1980)}]{Cordes80b}
{Cordes}, J.~M., \& {Helfand}, D.~J. 1980, \apj, 239, 640

\bibitem[{{D'Alessandro} {et~al.}(1995){D'Alessandro}, {McCulloch}, {Hamilton},
  \& {Deshpande}}]{DAlessandro95}
{D'Alessandro}, F., {McCulloch}, P.~M., {Hamilton}, P.~A., \& {Deshpande},
  A.~A. 1995, \mnras, 277, 1033

\bibitem[{{de Jager} {et~al.}(1989){de Jager}, {Raubenheimer}, \&
  {Swanepoel}}]{deJager89}
{de Jager}, O.~C., {Raubenheimer}, B.~C., \& {Swanepoel}, J.~W.~H. 1989, \aap,
  221, 180

\bibitem[{{De Luca} {et~al.}(2011){De Luca}, {Marelli}, {Mignani}, {Caraveo},
  {Hummel}, {Collins}, {Shearer}, {Saz Parkinson}, {Belfiore}, \&
  {Bignami}}]{DeLuca11}
{De Luca}, A., {Marelli}, M., {Mignani}, R.~P., {et~al.} 2011, \apj, 733, 104

\bibitem[{{Deller} {et~al.}(2008){Deller}, {Verbiest}, {Tingay}, \&
  {Bailes}}]{Deller08}
{Deller}, A.~T., {Verbiest}, J.~P.~W., {Tingay}, S.~J., \& {Bailes}, M. 2008,
  \apjl, 685, L67

\bibitem[{{Deng} {et~al.}(2012){Deng}, {Coles}, {Hobbs}, {Keith}, {Manchester},
  {Shannon}, \& {Zheng}}]{Deng12}
{Deng}, X.~P., {Coles}, W., {Hobbs}, G., {et~al.} 2012, \mnras, 424, 244

\bibitem[{{Dodson} {et~al.}(2003){Dodson}, {Legge}, {Reynolds}, \&
  {McCulloch}}]{Dodson03}
{Dodson}, R., {Legge}, D., {Reynolds}, J.~E., \& {McCulloch}, P.~M. 2003, \apj,
  596, 1137

\bibitem[{{Dormody} {et~al.}(2011){Dormody}, {Johnson}, {Atwood}, {Belfiore},
  {Grenier}, {Johnson}, {Razzano}, \& {Saz Parkinson}}]{Dormody11}
{Dormody}, M., {Johnson}, R.~P., {Atwood}, W.~B., {et~al.} 2011, \apj, 742, 126

\bibitem[{{Espinoza} {et~al.}(2014){Espinoza}, {Antonopoulou}, {Stappers},
  {Watts}, \& {Lyne}}]{Espinoza14}
{Espinoza}, C.~M., {Antonopoulou}, D., {Stappers}, B.~W., {Watts}, A., \&
  {Lyne}, A.~G. 2014, \mnras, 440, 2755

\bibitem[{{Espinoza} {et~al.}(2011){Espinoza}, {Lyne}, {Stappers}, \&
  {Kramer}}]{Espinoza11}
{Espinoza}, C.~M., {Lyne}, A.~G., {Stappers}, B.~W., \& {Kramer}, M. 2011,
  \mnras, 414, 1679

\bibitem[{{Faherty} {et~al.}(2007){Faherty}, {Walter}, \&
  {Anderson}}]{Faherty07}
{Faherty}, J., {Walter}, F.~M., \& {Anderson}, J. 2007, \apss, 308, 225

\bibitem[{{Foreman-Mackey} {et~al.}(2013){Foreman-Mackey}, {Hogg}, {Lang}, \&
  {Goodman}}]{Foreman-Mackey13}
{Foreman-Mackey}, D., {Hogg}, D.~W., {Lang}, D., \& {Goodman}, J. 2013, \pasp,
  125, 306

\bibitem[{{Gaensler} {et~al.}(2004){Gaensler}, {van der Swaluw}, {Camilo},
  {Kaspi}, {Baganoff}, {Yusef-Zadeh}, \& {Manchester}}]{Gaensler04}
{Gaensler}, B.~M., {van der Swaluw}, E., {Camilo}, F., {et~al.} 2004, \apj,
  616, 383

\bibitem[{{Gonzalez} {et~al.}(2006){Gonzalez}, {Kaspi}, {Pivovaroff}, \&
  {Gaensler}}]{Gonzalez06}
{Gonzalez}, M.~E., {Kaspi}, V.~M., {Pivovaroff}, M.~J., \& {Gaensler}, B.~M.
  2006, \apj, 652, 569

\bibitem[{{Halpern} {et~al.}(2001){Halpern}, {Camilo}, {Gotthelf}, {Helfand},
  {Kramer}, {Lyne}, {Leighly}, \& {Eracleous}}]{Halpern01}
{Halpern}, J.~P., {Camilo}, F., {Gotthelf}, E.~V., {et~al.} 2001, \apjl, 552,
  L125

\bibitem[{{Halpern} {et~al.}(2004){Halpern}, {Gotthelf}, {Camilo}, {Helfand},
  \& {Ransom}}]{Halpern04}
{Halpern}, J.~P., {Gotthelf}, E.~V., {Camilo}, F., {Helfand}, D.~J., \&
  {Ransom}, S.~M. 2004, \apj, 612, 398

\bibitem[{{Halpern} {et~al.}(2002){Halpern}, {Gotthelf}, {Mirabal}, \&
  {Camilo}}]{Halpern02}
{Halpern}, J.~P., {Gotthelf}, E.~V., {Mirabal}, N., \& {Camilo}, F. 2002,
  \apjl, 573, L41

\bibitem[{{Hermsen} {et~al.}(2013){Hermsen}, {Hessels}, {Kuiper}, {van
  Leeuwen}, {Mitra}, {de Plaa}, {Rankin}, {Stappers}, {Wright}, {Basu},
  {Alexov}, \& et~al.}]{Hermsen13}
{Hermsen}, W., {Hessels}, J.~W.~T., {Kuiper}, L., {et~al.} 2013, Science, 339,
  436

\bibitem[{{Hewish} {et~al.}(1968){Hewish}, {Bell}, {Pilkington}, {Scott}, \&
  {Collins}}]{Hewish68}
{Hewish}, A., {Bell}, S.~J., {Pilkington}, J.~D.~H., {Scott}, P.~F., \&
  {Collins}, R.~A. 1968, \nat, 217, 709

\bibitem[{{Hobbs} {et~al.}(2010{\natexlab{a}}){Hobbs}, {Lyne}, \&
  {Kramer}}]{Hobbs10b}
{Hobbs}, G., {Lyne}, A.~G., \& {Kramer}, M. 2010{\natexlab{a}}, \mnras, 402,
  1027

\bibitem[{{Hobbs} {et~al.}(2004){Hobbs}, {Lyne}, {Kramer}, {Martin}, \&
  {Jordan}}]{Hobbs04}
{Hobbs}, G., {Lyne}, A.~G., {Kramer}, M., {Martin}, C.~E., \& {Jordan}, C.
  2004, \mnras, 353, 1311

\bibitem[{{Hobbs} {et~al.}(2010{\natexlab{b}}){Hobbs}, {Archibald},
  {Arzoumanian}, {Backer}, {Bailes}, {Bhat}, {Burgay}, {Burke-Spolaor},
  {Champion}, {Cognard}, {Coles}, {Cordes}, {Demorest}, {Desvignes}, {Ferdman},
  {Finn}, {Freire}, {Gonzalez}, {Hessels}, {Hotan}, {Janssen}, {Jenet},
  {Jessner}, {Jordan}, {Kaspi}, {Kramer}, {Kondratiev}, {Lazio}, {Lazaridis},
  {Lee}, {Levin}, {Lommen}, {Lorimer}, {Lynch}, {Lyne}, {Manchester},
  {McLaughlin}, {Nice}, {Oslowski}, {Pilia}, {Possenti}, {Purver}, {Ransom},
  {Reynolds}, {Sanidas}, {Sarkissian}, {Sesana}, {Shannon}, {Siemens},
  {Stairs}, {Stappers}, {Stinebring}, {Theureau}, {van Haasteren}, {van
  Straten}, {Verbiest}, {Yardley}, \& {You}}]{Hobbs10}
{Hobbs}, G., {Archibald}, A., {Arzoumanian}, Z., {et~al.} 2010{\natexlab{b}},
  Classical and Quantum Gravity, 27, 084013

\bibitem[{{Hobbs} {et~al.}(2011){Hobbs}, {Miller}, {Manchester}, {Dempsey},
  {Chapman}, {Khoo}, {Applegate}, {Bailes}, {Bhat}, {Bridle}, {Borg}, {Brown},
  {Burnett}, {Camilo}, {Cattalini}, {Chaudhary}, {Chen}, {D'Amico},
  {Kedziora-Chudczer}, {Cornwell}, {George}, {Hampson}, {Hepburn}, {Jameson},
  {Keith}, {Kelly}, {Kosmynin}, {Lenc}, {Lorimer}, {Love}, {Lyne}, {McIntyre},
  {Morrissey}, {Pienaar}, {Reynolds}, {Ryder}, {Sarkissian}, {Stevenson},
  {Treloar}, {van Straten}, {Whiting}, \& {Wilson}}]{Hobbs11}
{Hobbs}, G., {Miller}, D., {Manchester}, R.~N., {et~al.} 2011, \pasa, 28, 202

\bibitem[{{Hobbs} {et~al.}(2006){Hobbs}, {Edwards}, \& {Manchester}}]{Hobbs06}
{Hobbs}, G.~B., {Edwards}, R.~T., \& {Manchester}, R.~N. 2006, \mnras, 369, 655

\bibitem[{{Hotan} {et~al.}(2005){Hotan}, {Bailes}, \& {Ord}}]{Hotan05}
{Hotan}, A.~W., {Bailes}, M., \& {Ord}, S.~M. 2005, \mnras, 362, 1267

\bibitem[{{Hughes} {et~al.}(2003){Hughes}, {Slane}, {Park}, {Roming}, \&
  {Burrows}}]{Hughes03}
{Hughes}, J.~P., {Slane}, P.~O., {Park}, S., {Roming}, P.~W.~A., \& {Burrows},
  D.~N. 2003, \apjl, 591, L139

\bibitem[{{Johnson} {et~al.}(2013){Johnson}, {Guillemot}, {Kerr}, {Cognard},
  {Ray}, {Wolff}, {B{\'e}gin}, {Janssen}, {Romani}, {Venter}, {Grove},
  {Freire}, {Wood}, {Cheung}, {Casandjian}, {Stairs}, {Camilo}, {Espinoza},
  {Ferrara}, {Harding}, {Johnston}, {Kramer}, {Lyne}, {Michelson}, {Ransom},
  {Shannon}, {Smith}, {Stappers}, {Theureau}, \& {Thorsett}}]{Johnson13}
{Johnson}, T.~J., {Guillemot}, L., {Kerr}, M., {et~al.} 2013, \apj, 778, 106

\bibitem[{{Kaplan} {et~al.}(2008){Kaplan}, {Chatterjee}, {Gaensler}, \&
  {Anderson}}]{Kaplan08}
{Kaplan}, D.~L., {Chatterjee}, S., {Gaensler}, B.~M., \& {Anderson}, J. 2008,
  \apj, 677, 1201

\bibitem[{{Kargaltsev} {et~al.}(2012){Kargaltsev}, {Durant}, {Pavlov}, \&
  {Garmire}}]{Kargaltsev12}
{Kargaltsev}, O., {Durant}, M., {Pavlov}, G.~G., \& {Garmire}, G. 2012, \apjs,
  201, 37

\bibitem[{{Kargaltsev} {et~al.}(2008){Kargaltsev}, {Misanovic}, {Pavlov},
  {Wong}, \& {Garmire}}]{Kargaltsev08}
{Kargaltsev}, O., {Misanovic}, Z., {Pavlov}, G.~G., {Wong}, J.~A., \&
  {Garmire}, G.~P. 2008, \apj, 684, 542

\bibitem[{{Keith} {et~al.}(2008){Keith}, {Johnston}, {Kramer}, {Weltevrede},
  {Watters}, \& {Stappers}}]{Keith08}
{Keith}, M.~J., {Johnston}, S., {Kramer}, M., {et~al.} 2008, \mnras, 389, 1881

\bibitem[{{Kerr}(2011)}]{Kerr11}
{Kerr}, M. 2011, \apj, 732, 38

\bibitem[{{Kerr} {et~al.}(2012){Kerr}, {Camilo}, {Johnson}, {Ferrara},
  {Guillemot}, {Harding}, {Hessels}, {Johnston}, {Keith}, {Kramer}, {Ransom},
  {Ray}, {Reynolds}, {Sarkissian}, \& {Wood}}]{Kerr12}
{Kerr}, M., {Camilo}, F., {Johnson}, T.~J., {et~al.} 2012, \apjl, 748, L2

\bibitem[{{Kramer} {et~al.}(2006){Kramer}, {Lyne}, {O'Brien}, {Jordan}, \&
  {Lorimer}}]{Kramer06}
{Kramer}, M., {Lyne}, A.~G., {O'Brien}, J.~T., {Jordan}, C.~A., \& {Lorimer},
  D.~R. 2006, Science, 312, 549

\bibitem[{{Lyne} {et~al.}(2010){Lyne}, {Hobbs}, {Kramer}, {Stairs}, \&
  {Stappers}}]{Lyne10}
{Lyne}, A., {Hobbs}, G., {Kramer}, M., {Stairs}, I., \& {Stappers}, B. 2010,
  Science, 329, 408

\bibitem[{{Marelli}(2012)}]{Marelli12}
{Marelli}, M. 2012, ArXiv e-prints, arXiv:1205.1748

\bibitem[{{Mignani} {et~al.}(2010){Mignani}, {Pavlov}, \&
  {Kargaltsev}}]{Mignani10}
{Mignani}, R.~P., {Pavlov}, G.~G., \& {Kargaltsev}, O. 2010, \apj, 720, 1635

\bibitem[{{Ng} {et~al.}(2005){Ng}, {Roberts}, \& {Romani}}]{Ng05}
{Ng}, C.-Y., {Roberts}, M.~S.~E., \& {Romani}, R.~W. 2005, \apj, 627, 904

\bibitem[{{Nolan} {et~al.}(2012){Nolan}, {Abdo}, {Ackermann}, {Ajello},
  {Allafort}, {Antolini}, {Atwood}, {Axelsson}, {Baldini}, {Ballet}, \&
  et~al.}]{Nolan12}
{Nolan}, P.~L., {Abdo}, A.~A., {Ackermann}, M., {et~al.} 2012, \apjs, 199, 31

\bibitem[{{Perrodin} {et~al.}(2013){Perrodin}, {Jenet}, {Lommen}, {Finn},
  {Demorest}, {Ferdman}, {Gonzalez}, {Nice}, {Ransom}, \&
  {Stairs}}]{Perrodin13}
{Perrodin}, D., {Jenet}, F., {Lommen}, A., {et~al.} 2013, ArXiv e-prints,
  arXiv:1311.3693

\bibitem[{{Pletsch} {et~al.}(2013){Pletsch}, {Guillemot}, {Allen}, {Anderson},
  {Aulbert}, {Bock}, {Champion}, {Eggenstein}, {Fehrmann}, {Hammer},
  {Karuppusamy}, {Keith}, {Kramer}, {Machenschalk}, {Ng}, {Papa}, {Ray}, \&
  {Siemens}}]{Pletsch13}
{Pletsch}, H.~J., {Guillemot}, L., {Allen}, B., {et~al.} 2013, \apjl, 779, L11

\bibitem[{{Ransom} {et~al.}(2011){Ransom}, {Ray}, {Camilo}, {Roberts}, {{\c
  C}elik}, {Wolff}, {Cheung}, {Kerr}, {Pennucci}, {DeCesar}, {Cognard}, {Lyne},
  {Stappers}, {Freire}, {Grove}, {Abdo}, {Desvignes}, {Donato}, {Ferrara},
  {Gehrels}, {Guillemot}, {Gwon}, {Harding}, {Johnston}, {Keith}, {Kramer},
  {Michelson}, {Parent}, {Saz Parkinson}, {Romani}, {Smith}, {Theureau},
  {Thompson}, {Weltevrede}, {Wood}, \& {Ziegler}}]{Ransom11}
{Ransom}, S.~M., {Ray}, P.~S., {Camilo}, F., {et~al.} 2011, \apjl, 727, L16

\bibitem[{{Ray} {et~al.}(2011){Ray}, {Kerr}, {Parent}, {Abdo}, {Guillemot},
  {Ransom}, {Rea}, {Wolff}, {Makeev}, {Roberts}, {Camilo}, {Dormody}, {Freire},
  {Grove}, {Gwon}, {Harding}, {Johnston}, {Keith}, {Kramer}, {Michelson},
  {Romani}, {Saz Parkinson}, {Thompson}, {Weltevrede}, {Wood}, \&
  {Ziegler}}]{Ray11}
{Ray}, P.~S., {Kerr}, M., {Parent}, D., {et~al.} 2011, \apjs, 194, 17

\bibitem[{{Romani} {et~al.}(2010){Romani}, {Shaw}, {Camilo}, {Cotter}, \&
  {Sivakoff}}]{Romani10}
{Romani}, R.~W., {Shaw}, M.~S., {Camilo}, F., {Cotter}, G., \& {Sivakoff},
  G.~R. 2010, \apj, 724, 908

\bibitem[{{Shannon} \& {Cordes}(2010)}]{Shannon10}
{Shannon}, R.~M., \& {Cordes}, J.~M. 2010, \apj, 725, 1607

\bibitem[{{Slane} {et~al.}(2002){Slane}, {Helfand}, \& {Murray}}]{Slane02}
{Slane}, P.~O., {Helfand}, D.~J., \& {Murray}, S.~S. 2002, \apjl, 571, L45

\bibitem[{{Smith} {et~al.}(2008){Smith}, {Guillemot}, {Camilo}, {Cognard},
  {Dumora}, {Espinoza}, {Freire}, {Gotthelf}, {Harding}, {Hobbs}, {Johnston},
  {Kaspi}, {Kramer}, {Livingstone}, {Lyne}, {Manchester}, {Marshall},
  {McLaughlin}, {Noutsos}, {Ransom}, {Roberts}, {Romani}, {Stappers},
  {Theureau}, {Thompson}, {Thorsett}, {Wang}, \& {Weltevrede}}]{Smith08}
{Smith}, D.~A., {Guillemot}, L., {Camilo}, F., {et~al.} 2008, \aap, 492, 923

\bibitem[{{Stappers} {et~al.}(1999){Stappers}, {Gaensler}, \&
  {Johnston}}]{Stappers99}
{Stappers}, B.~W., {Gaensler}, B.~M., \& {Johnston}, S. 1999, \mnras, 308, 609

\bibitem[{{Tegmark} {et~al.}(1997){Tegmark}, {Taylor}, \&
  {Heavens}}]{Tegmark97}
{Tegmark}, M., {Taylor}, A.~N., \& {Heavens}, A.~F. 1997, \apj, 480, 22

\bibitem[{{Van Etten} {et~al.}(2008){Van Etten}, {Romani}, \&
  {Ng}}]{VanEtten08}
{Van Etten}, A., {Romani}, R.~W., \& {Ng}, C.-Y. 2008, \apj, 680, 1417

\bibitem[{{Van Etten} {et~al.}(2012){Van Etten}, {Romani}, \&
  {Ng}}]{VanEtten12}
---. 2012, \apj, 755, 151

\bibitem[{{van Haasteren} \& {Levin}(2013)}]{VanHaasteren13}
{van Haasteren}, R., \& {Levin}, Y. 2013, \mnras, 428, 1147

\bibitem[{{Watters} \& {Romani}(2011)}]{Watters11}
{Watters}, K.~P., \& {Romani}, R.~W. 2011, \apj, 727, 123

\bibitem[{{Weisskopf} {et~al.}(2011){Weisskopf}, {Romani}, {Razzano},
  {Belfiore}, {Saz Parkinson}, {Ray}, {Kerr}, {Harding}, {Swartz},
  {Carrami{\~n}ana}, {Ziegler}, {Becker}, {De Luca}, {Dormody}, {Thompson},
  {Kanbach}, {Elsner}, {O'Dell}, \& {Tennant}}]{Weisskopf11}
{Weisskopf}, M.~C., {Romani}, R.~W., {Razzano}, M., {et~al.} 2011, \apj, 743,
  74

\bibitem[{{Weltevrede} {et~al.}(2010){Weltevrede}, {Johnston}, {Manchester},
  {Bhat}, {Burgay}, {Champion}, {Hobbs}, {K{\i}z{\i}ltan}, {Keith}, {Possenti},
  {Reynolds}, \& {Watters}}]{Weltevrede10}
{Weltevrede}, P., {Johnston}, S., {Manchester}, R.~N., {et~al.} 2010, \pasa,
  27, 64

\bibitem[{{Zeiger} {et~al.}(2008){Zeiger}, {Brisken}, {Chatterjee}, \&
  {Goss}}]{Zeiger08}
{Zeiger}, B.~R., {Brisken}, W.~F., {Chatterjee}, S., \& {Goss}, W.~M. 2008,
  \apj, 674, 271

\end{thebibliography}


\end{document}